# Tetragonal CH₃NH₃PbI₃ Is Ferroelectric

Yevgeny Rakita[1], Omri Bar-Elli[2], Elena Meirzadeh[1], Hadar Kaslasi[1], Yagel Peleg[1], Gary Hodes[1], Igor Lubomirsky[1], Dan Oron[2], David Ehre[1,a], David Cahen[1,a]

[1] *Materials and Interfaces Dep., Weizmann Institute of Science, 7610001, Rehovot, Israel*
[2] *Physics of Complex Systems Dep., Weizmann Institute of Science, 7610001, Rehovot, Israel*
[a] *Electronic mail: david.ehre@weizmann.ac.il ; david.cahen@weizmann.ac.il*

**Abstract:**

Halide perovskite (HaP) semiconductors are revolutionizing photovoltaic (PV) solar energy conversion by showing remarkable performance of solar cells made with *esp*. tetragonal methylammonium lead tri-iodide (MAPbI₃). In particular, the low voltage loss of these cells implies a remarkably low recombination rate of photogenerated carriers. It was suggested that low recombination can be due to spatial separation of electrons and holes, a possibility if MAPbI₃ is a semiconducting *ferroelectric*, which, however, requires clear experimental evidence. As a first step we show that, *in operando*, MAPbI₃ (unlike MAPbBr₃) is pyroelectric, which implies it *can* be ferroelectric. The next step, proving it *is* (not) ferroelectric, is challenging, because of the material's relatively high electrical conductance (a consequence of an optical band gap suitable for PV conversion!) and low stability under high applied bias voltage. This excludes normal measurements of a ferroelectric hysteresis loop to prove ferroelctricity's hallmark for switchable polarization. By adopting an approach suitable for electrically leaky materials as MAPbI₃, we show here ferroelectric hysteresis from well-characterized single crystals at low temperature (still within the tetragonal phase, which is the room temperature stable phase). Using chemical etching, we also image polar domains, the structural fingerprint for ferroelectricity, periodically stacked along the polar axis of the crystal, which, as predicted by theory, scale with the overall crystal size. We also succeeded in detecting clear second-harmonic generation, direct evidence for the material's non-centrosymmetry. We note that the material's ferroelectric nature, can, but not obviously need to be important in a PV cell, operating around *room temperature*.

\* *Supporting information is after the Reference section.*

**Introduction**

New optoelectronic materials are of interest for solar cells with higher power and voltage efficiencies, lower costs and improved long-term reliability. A very recent entry is the family of halide perovskites (HaPs), in particular those based on methylammonium (MA) lead iodide (MAPbI$_3$), MAPbBr$_3$ and its inorganic analog CsPbBr$_3$. Devices based on these, perform remarkably well as solar cells (1, 2), as well as for other optoelectronic applications, such as LEDs and electromagnetic radiation detectors (3–5). Understanding possible unique characteristics of HaPs may show the way to other materials with similar key features.

The *ABX$_3$* (X=I, Br, Cl) halide perovskite semiconductors (SC), *i.e.*, with perovskite or perovskite-like structures, reach, *via* a steep absorption edge, a high optical absorption coefficient ($\sim 10^5$ cm$^{-1}$) (6, 7), long charge carrier lifetimes ($\sim 0.1$-1 μs) (8), reasonable carrier mobilities ($\leq \sim 100$ cm$^2 \cdot$V$^{-1} \cdot$s$^{-1}$) (9) and have a low exciton binding energy (10). With these characteristics the width of the optical absorber layer can be $\leq 0.5$ μm, which allows the charge carriers (separated electrons and holes) to diffuse/drift throughout the entire width of the absorber without strong recombination. HaP-based solar-cell devices can, usually, function with a continuous HaP layer, *i.e.*, also without filling a high surface area, non-HaP, scaffold, such as mp-TiO$_2$ (11). Photogenerated electrons and holes will then spend a significant time in the bulk of the HaP film. Consequently, because the dielectric nature of a SC strongly affects the charge dynamics in it, the performance of HaP-based devices should be strongly affected by the dielectric behaviour of the HaP itself. Here we present experimental results that bear on possible ferroelectricity, the type of dielectric behavior that has been suggested to aid photovoltaic activity of HaPs, in particular MAPbI$_3$, which is one of the materials that has yielded high efficiency solar cells.

*Ferroelectricity* is the ability to change the spontaneous polarization in a material by an external electric field. It is a well-known dielectric phenomenon, existing in many oxide perovskites (12, 13), and has been suggested to be present in HaPs, in particular MAPbI$_3$ (14, 15). A short overview about '*Origins of polarity in perovskite and its detection*' is given in Section 1 in the Supporting Information. If the material is indeed ferroelectric then this suggests the possible existence of bulk photovoltaics, which can arise due to existing spontaneous polarization in the bulk of the material (16). Ferroelectric MAPbI$_3$ could, however, have as main benefit (as broadly discussed elsewhere) (14–17) charge transport via *domain walls* between adjacent *polar* domains. The existence of polar domains induces charge separation, lowers charge recombination and allows high conductivity due to local degeneracy of the SC along the domain walls (18, 19). If this is so, then it can explain the remarkably high voltage efficiency, *i.e.*, low voltage loss in HaP-based solar cells (estimated as the difference between, or ratio of open-circuit voltage and band gap, ($E_G$-$V_{OC}$) and ($V_{OC}/E_G$), respectively, (20, 21) *esp.* for those with Iodide as the halide. A further indication for a low recombination rate of photogenerated carriers is the respectable voltage and current at maximum power that these HaP-based devices possess, which makes the question if HaPs can be ferroelectric of more than fundamental scientific interest.

Whereas existence of ferroelectricity in MAPbI$_3$ was suggested *theoretically* (14, 15), evidence on whether HaPs actually possess the domain structure that typifies a ferroelectric material are found to be quite contradictory (21–28). To analyze the question if HaP are *actually* ferroelectric, we need to test for the presence/absence of several fundamental phenomena, which are tightly related to it.

Necessary conditions for bulk ferroelectricity to exist are as follows:

*(a)* absence of an inversion symmetry (*i.e.* a **non-centrosymmetric** material), which may lead to

*(b)* a unit-cell with a permanent **polarity**, where

*(c)* an assembly of polar unit cells facing the same direction, which will form a so-called *'polar-domain'*. An assembly of periodically ordered polar domains, facing different directions (usually 180° or 90° to each other), will form the bulk of a ferroelectric material.

*(d)* The polarity of these polar domains must switch when sufficiently high electric fields are applied.

The most straightforward way to prove a material is ferroelectric is by showing switchable polarization ($P$) under an applied electric field ($E$), which results in a commonly presented $P$-$E$ hysteresis loop (29, 30). In materials for which the energetic cost to reverse their polarity is too high (due to lattice rigidity, as is the case for ZnO) or the required electric potential is higher than the dissociation potential of the material, observing a switchable polarization may be very challenging. In MAPbI$_3$, which has a low bandgap (1.55-1.60 eV) (31) with relatively high electronic, and possibly some ionic conductivity (32), with low formation energy (33, 34), detection of possible ferroelectric switching is challenging. Indeed, previous attempts to find such evidence (22, 23) showed contradicting results to what was predicted from theory (14, 15), Here we use the *lossy* part of the dielectric response (in contrast to using the *energy storage* part of the dielectric response as can be done with insulating ferroelectrics, *cf.* Fig. 1), which yields clear ferroelectric $P$-$E$ behaviour from tetragonal MAPbI$_3$ crystals.

Still, as this is a highly debated issue, further evidence for ferroelectricity of the material must exist and is, therefore, shown. Under steady-state conditions (*i.e.* under zero applied field) there will be an energetic cost to having a dipole in the bulk of a solid. One (or more) of the following screening mechanisms will, therefore, be active (18):

a)      **Accumulation of oppositely charged species at the outer surface of the polar bulk** (*e.g.*, adsorption of charged molecules or free charges from the environment or adjacent electrodes, respectively). By depositing electrodes on opposite crystal faces, one can measure the reorganization charge flow of these charges, when the temperature, which changes the spontaneous polarization, is changed. This phenomenon is called '*pyroelectricity*' and is direct evidence for existing polarity. Earlier we showed clear experimental evidence that at RT cubic MAPbBr$_3$ is not pyroelectric (21). Here we prove experimentally that tetragonal MAPbI$_3$, the stable RT phase of this material, is pyroelectric and, thus, polar.

b)      **Accumulation of mobile electrons and holes from the polar bulk material itself** can occur if the bulk materials is sufficiently conductive, $> \sim 10^{-6}$ S·cm$^{-1}$, a reasonable value for a low bandgap SC, such as MAPbI$_3$. In this sense, leakage currents should dominate the dielectric response and, as we will see, are affected by the bulk overall polarization.

c)      **Stacking of domains of opposite polarity to decrease Coulomb repulsion, but at the cost of forming a 'domain-wall'.** For many known ferroelectric (as well as ferromagnetic) materials, the periodicity between the domains, $\omega$, scales with the crystal thickness along its polar axis, $D$, as $\omega \sim D^{\gamma}$. '$\gamma$' is the power dependence that theoretically should be equal to 0.5 (according to the Landau-Lifshitz-Kittel model) (35, 36). We show that the tetragonal-MAPbI$_3$ follows this rule.

To complete the picture, as polar materials are also non-centrosymmetric, we show positive results of second-harmonic-generation (SHG), which must exist in a polar medium (37). Based on our results related to domain size and orientation we explain why previous attempts to detect SHG in MAPbI$_3$ were not successful (24).

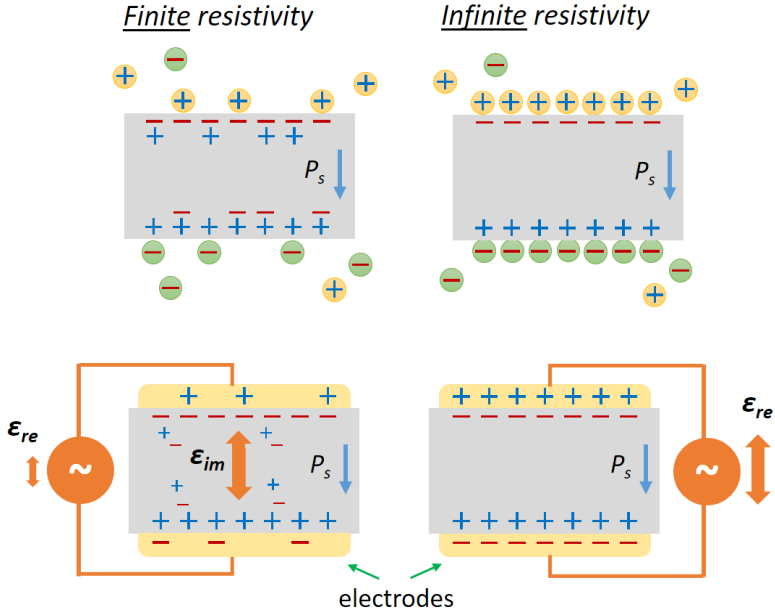

**Fig. 1**: *Illustration of the conceptual difference between (a) a lossy SC and (b) an insulator*; both are shown without (top) and with (bottom) electrodes. Under an applied AC bias, for the lossy SC the dielectric response related to energy dissipation ($\varepsilon_{im}$) will dominate, while for the insulator the dielectric response related to energy storage ($\varepsilon_{re}$) will dominate. '$P_s$' represents a spontaneous polarization.

## Results and Discussion

### *Pyroelectricity*

As in our earlier work on MAPbBr$_3$ (21), we used periodic pulse pyroelectric measurements (41–43) to determine if tetragonal MAPbI$_3$ is polar along its $c$ axis. Pyroelectric currents can be distinguished from other possible thermally stimulated electric response (TSER) (thermoelectric or flexoelectricity) (21, 44) by heating once one, and once the other side of the crystal periodically. Pyroelectric currents will reverse their sign once a crystal, together with its electrical leads, is flipped (by 180° on its other face). Thermoelectricity, though, is generated due to a temperature gradient between the two sides of a (semi-) conducting bulk, and will not depend on an internal direction of a dipole. Therefore, two complementary measurements were done, heating once one, and once the other side of a crystal.

Fig. 2(a) shows that along the <001> direction (if the electrodes are deposited on the {001} planes) a clear pyroelectric response, $J_{pyro}$, is observed. The experiment is done at low temperatures (LT), ~ 210 K, to reduce any thermoelectric contribution to the TSER. Direct proportionality between a semiconductor's thermoelectricity and conductivity imply that thermoelectric currents will become more dominant at higher temperatures, because the conductivity of a semiconductor increases with temperature (see also Fig. S2(c)). Although at RT the pyroelectric response is still present, it is much weaker that at LT (see Fig. S2(a) and (b)), where it become the dominating part of the TSER. Apart from reduced thermoelectricity, lower electrical conductivity also should decrease leakage currents and, usually increase the effective spontaneous polarization. Both of these two factors should increase the pyroelectric response.

Fig. 2(b) shows, based on the peak-current value dependence on temperature, further evidence for the pyroelectric nature of MAPbI$_3$. The local maximum around the phase transition temperature ($T_c$) at 330 K reflects a rise in the pyroelectric response as expected from Ginsburg-Landau theory (45). Thus, a local extremum around the phase transition temperature (330 K), due to a discontinuous change in the dipole ($J_{pyro} \propto \Delta P$) at a polar to non-polar phase transition, *i.e.*, around the $T_c$, tetragonal to cubic, in this case, should occur. The disappearance and reappearance of the pyroelectric response after exceeding the $T_c$ to the cubic phase and cooling back to the tetragonal phase, lends further support to the polar nature of tetragonal MAPbI$_3$ and the non-polar nature of the cubic phase, a result that agrees with our results for MAPbBr$_3$ (21). This also indicates that the response to the heat pulse is not related to trapped charges, as commonly found for MAPbBr$_3$.

The effect of the growth environment, which may induce a permanent dipole due to defects (*e.g.*, as result of the presence of anti-sol or stabilizer) (46), is checked by measuring pyroelectricity from crystals grown in different environments, all below $T_c$ (see Fig. S2(a) and (b)). Although there are some differences in the TSER profile, the basic pyroelectric behaviour persists in all cases, a strong indication that pure tetragonal MAPbI$_3$ is polar, with a *I4cm* space group (and not *I4/mcm* as occasionally reported (24)). Furthermore, the fast rise, followed by a decay, as seen in the TSER profile for all three types of crystals, indicates a polar-domain-like structure, as expected from a ferroelectric material. Further proof for ferroelectricity is presented in the following sections.

We note that the current measurements were carried out only after the surrounding temperature around the crystal was stabilized to ±1 K. This is done to avoid currents that are generated by a flexoelectric polarization (44, 47, 48), as can occur upon fast heating or cooling (see Fig. S4). Measuring along the non-polar {110} direction under stable temperature conditions, did not show any evidence of pyroelectricity. This is expected for measurement along a non-polar direction where the polarization (which equals zero) does not change upon pulsed heating.

To check if other <u>*tetragonal*</u> MAPbX$_3$ compounds are also pyroelectric, we measured *tetragonal* MAPbBr$_3$ (well below its $T_c$, which = 235 K) (49) and found no pyroelectric response (Fig. S5). Since the crystals were grown in their cubic phase, the absence of a pyroelectric response in *tetragonal* MAPbBr$_3$ may be due to the absence of polar domains, or due to formation of randomly oriented polar domains that create an overall absence of polarization. Distinguishing between these options will require further investigation.

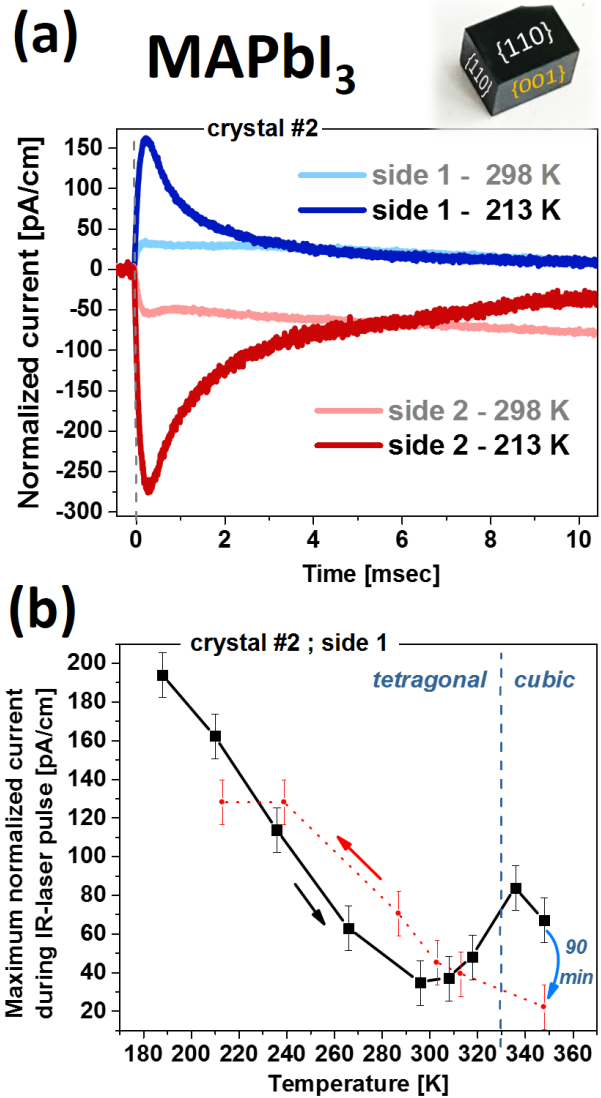

**Fig. 2**: *Pyroelectric response measurements*; *(a) Pyroelectric response at RT (pale-colored plots) and 213 K, LT (deep-colored plots), with contacts on the [001] planes of a MAPbI$_3$ crystal. At RT the response is convoluted with some thermoelectric currents. Evidence from different crystals can be found in Fig. S2. (b) Peak-current value dependence on temperature, which shows a TSER that is characteristic for a pyroelectric material (for details, see main text). Evidence for all the points in the plot can be found in Fig. S3.*

## *Ferroelectricity*

Classic, insulating, ferroelectric materials, where leakage currents are negligible, commonly show a distinctive dependence of their relative permittivity, $\varepsilon$, on a bias electric field, $E$, (see Fig. S6(a)) (30, 50). For such insulating materials, the real part of the relative permittivity, $\varepsilon_{re}$, which is the dielectric response related to energy storage, dominates the complex relative permittivity. The imaginary part of the relative permittivity, $\varepsilon_{im}$ (the dielectric response related to energy dissipation), is, therefore, usually neglected (see Fig. S6(b) and Fig. 1(b)).

In *semiconducting* ferroelectric materials, however, leakage currents, which are represented by $\varepsilon_{im}$ (see Fig. 1(a)) can become a significant part of the response (23, 51), as shown in Fig. 3(a). To reconstruct a *P-E* hysteresis loop we can simply integrate $\varepsilon_{im}$ over the bias electric field, $E_{DC}$, based on the following equation:

1)      $\Delta P = \varepsilon_0 \int \varepsilon \cdot \partial E \approx \varepsilon_0 \int \varepsilon_{im} \cdot \partial E$ .

Here $\Delta P$ is the change in polarization from the initial one and $\varepsilon_0$ is the vacuum permittivity. The first part of the equation is the relation between $P$ and $E$, where the dielectric constant, which is directly related to the susceptibility, is a proportionality factor. The last part results directly for the case where $\varepsilon_{re}$ can be neglected.

Fig. 3(b) shows the result of the integration of $\varepsilon_{im}$ over $E_{DC}$ measured on a tetragonal MAPbI$_3$ crystal. The shape of the hysteresis loop resembles clearly a ferroelectric polarization as function of applied bias. MAPbI$_3$ was claimed to be ferroelectric before, but our result presents the first clear evidence for ferroelectric *P-E* switching in single crystals of tetragonal MAPbI$_3$. This result is a direct outcome from using the dissipative part of the permittivity to observe polarization switching of a ferroelectric bulk. In the previous cases, only the real part of the permittivity was similarly analyzed (22, 23, 52).

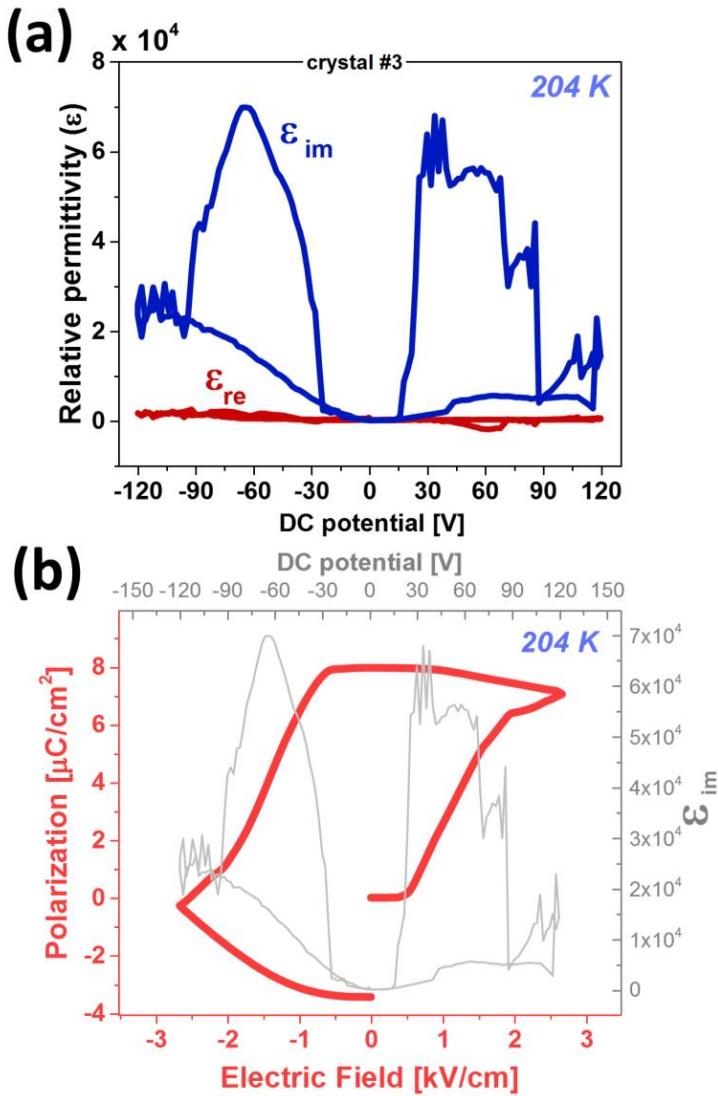

**Fig. 3**: *Ferroelectric response measurement; (a) Dielectric response at 204 K of a MAPbI$_3$ crystal along its <001> direction, measured at 2,222 Hz and V$_{AC}$ =0.1 V, as a function of applied bias, E$_{DC}$ (X-axis), showing that the imaginary (dissipative) part dominates the dielectric response. (b) P-E hysteresis loop obtained from integration of $\varepsilon_{im}$ over E$_{DC}$ (see Eq. 1). The hysteresis loop is convoluted from a lossy bulk (i.e., resistor) response and ferroelectric polarization response, as illustrated in Fig. S7(a).*

In the case of MAPbI$_3$, the $\varepsilon_{im} - E$ dependence should be measured at as low a temperature as possible (in our case at 204 K, to be well above the next tetragonal$\leftrightarrow$ orthorhombic phase transition) for the following reasons: (a) the spontaneous polarization should increase with decreasing temperature, following the pyroelectric measurements (*cf.* Fig. 2(b)); (b) The rate of electrochemical and other decomposition reactions decreases with reduced temperature and conductivity; (c) ionic conduction and/or dielectric relaxation, which can screen the ferroelectric behavior, decrease with decreasing temperature. The last possible reason (c) is consistent with the frequency dependence of $\varepsilon_{re}$ (Fig. 4), with a much larger increase in $\varepsilon_{re}$ with decreasing frequency at RT than at 203 K. The reconstructed *P-E* hysteresis loop has the characteristics of a semiconducting ferroelectric material (53), the above-noted convolution of a ferroelectric part with a resistor part (see Fig. S7(a)).

The measured coercive field, $E_c$, is ~ 0.6 kV·cm$^{-1}$, and the saturation field, $E_{sat}$, is around ~ 2 kV·cm$^{-1}$ (see Fig. S7(b)). The saturation change in polarization, $\Delta P_{sat}$, is ~ 7 μC·cm$^{-2}$, which is very close to 8 μC·cm$^{-2}$ predicted by Fan *et al.*(23) for MAPbI$_3$ at RT, while somewhat smaller than the 38 μC·cm$^{-2}$ predicted by Frost *et al.* (54).

At RT, the sample decomposed around -1.2 kV·cm$^{-1}$ (*cf.* Fig. S8). Before decomposition occurs, the *P-E* loop resembles a ferroelectric loop, although such loop can originate also from electrochemical or ionic migration effects (55), which are not negligible at 2,222 Hz as can be seen in Fig. 4.

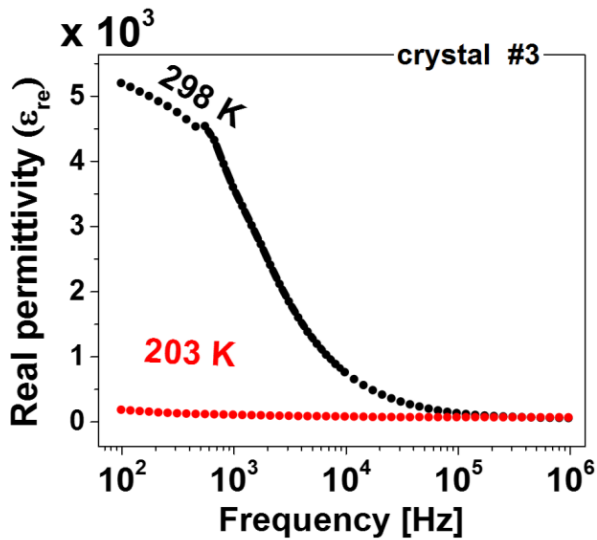

**Fig. 4**: *Relative permittivity dependence on the sweeping frequency at RT and LT. The RT response shows a sharp increase with decreasing frequency, starting above 0.1 MHz, which can be related to ionic migration or dielectric relaxation response to an applied field (0.1 $V_{AC}$ in this case).*

### *Ferroelectric domain-size*:

After showing MAPbI$_3$ responds to an electric bias as a ferroelectric material, and to periodic thermal change, it is now left to prove the existence of polar domains to complete the picture. Due to a different charge density between adjacent domains and/or between a domain and a domain-wall, the chemical activity should vary between the different parts. Following this logic, chemical etching should uncover a periodic structuring as was demonstrated for different ferroelectric materials (56–58).

After testing several solvents, acetone was found to reveal beautifully ordered periodic domains stacked along the polar axis of the crystal (see Fig. 5(a) and S10). This exposed pattern can explain both the smooth cleaving along the *c* (polar) planes of the crystal and the jagged cleaving perpendicular to it that we find. This simple etching procedure can also reveal twin grain boundaries at which the polar domains are perpendicular to each other (Fig. S9 (b)).

Another indication for the existence of polar domains is the decaying TSER profile. When polar domains are stacked opposite to each other (head-to-head, 180°) the TSER should vanish when the externally induced temperature change is equilibrated between two adjacent domains. The rate of this equilibration is related to the polar domain thickness, or domain period – $\omega$, and the thermal diffusivity, $\delta$, of the polar bulk. Taking the 200 K value of $\delta = 6 \times 10^{-3}$ cm$^2$·s$^{-1}$ for MAPbI$_3$ (59), and the decay time from the maximum current till it reaches its half-maximum value (subtracting the system delay – 0.02 ms), $t_{HM}$, one can estimate $\omega$ by using the following equation (for further information see Section 2 in the Supporting Information) (60):

1)  $\omega = 0.48\sqrt{4 \cdot \delta \cdot t_{HM}}$ .

With $t_{HM} = 0.16$ ms, we find that $\omega \sim 9$ μm, which is comparable, in terms of order of magnitude, to what we find from chemical etching (see Fig. 5(a) and S10).

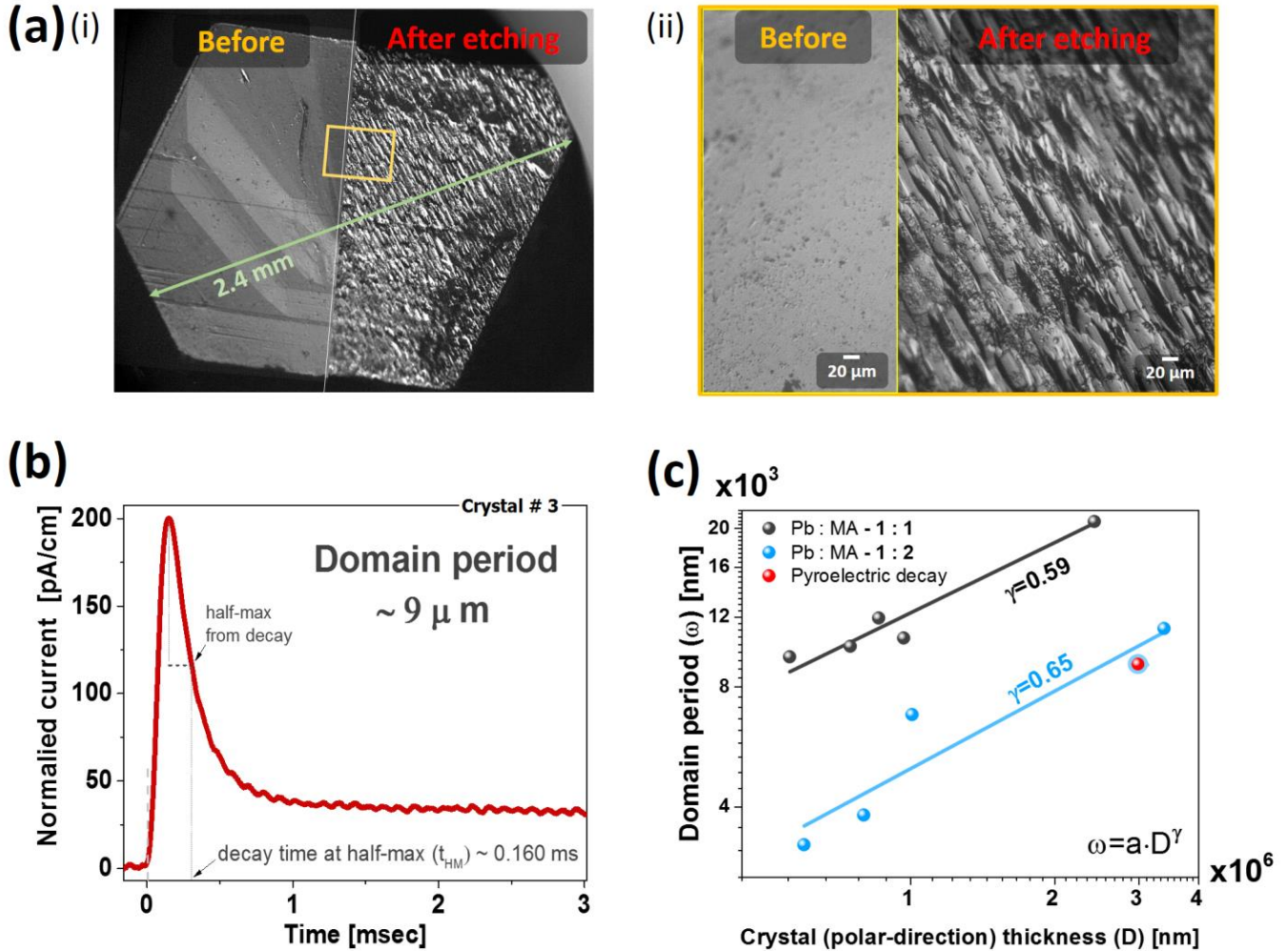

*Fig. 5: Evidence for polar domains and their scalability with crystal size; (a) Bright field image from a light microscope of a crystal before and after etching in acetone for 120 sec. The crystal (2.4 mm) was grown from a 1:1 PbI$_2$:MAI solution, using ethyl-acetate as anti-solvent (for comparison, an example for 0.56 mm crystal is shown in Fig. S9). (ii) Magnification of the yellow rectangle marked in (i). (b) The pyroelectric decay from crystal #3 (Fig. S2(b)) when measured with amplification of 10$^7$ (discharge time ~ 0.02 msec). Crystal #3 was grown from a 1:2 PbI$_2$ to MAI solution. Based on the analysis related to Eq. 2 the domain period should be ~ 9 μm. (c) Domain periodicity with respect to the crystal thickness in the polar direction. The data are an average derived from microscope images after etching in acetone for ~ 120 sec. The red dot represents the value obtained from the pyroelectric TSER (crystal #3) from (b). The values of γ are obtained by fitting to ω=a·D$^γ$.*

$\omega$ should scale with the total crystal thickness along its polar axis, $D$, as $\omega = a \cdot D^{\gamma}$, where '$a$' is a pre-factor that depends on the domain-wall thickness and '$\gamma$' is theoretically = 0.5 (35, 36) and empirically between $\sim$ 0.4 and 0.6 (61–63). By relating $\omega$ to $D$ from several single crystals (without grain boundaries; *cf.* Fig. S9(b)), in Fig. 5(c) we find $\omega \sim D^{0.6}$. Despite an uncertainty in $\gamma$ due to having only a few points over a small range of crystal sizes, it is interesting to note that for BiFeO$_3$, which is a known ferroelectric SC, $\gamma$ was found to be 0.59. Comparison with scaling relations of other ferroic materials is shown in Fig. S10. It is also found that crystals grown in 1:1 or 1:2 PbI$_2$ to MAI solutions give a slightly different result, mostly in the $a$ pre-factor. Since an excess of MAI was found to reduce carrier concentration in MAPbI$_3$ films (64), this small difference may be related to a doping effect.

Although the scaling relation is clear, quantitative extrapolation to dimensions of thin films will require further tuning. For instance, by extrapolating to $D$=100 nm, $\omega$ of the 1:1 and 1:2 (of PbI$_2$:MAI) cases result in domains of 56 or 13 nm, respectively. For the former case it would mean less than 2 domains in the sample in the polar direction. In the latter it implies that each domain contains $< \sim$ 10 unit cells (40) – a value that should be treated with caution. On the other hand, on the scale where $D$=2 $\mu$m, the typical size of a crystal shown in Hermens *et al.* (26), $\omega$ reaches values of few 100s of nm, which is of the scale of the ferro-*elastic* domains they observed.

### *Second Harmonic Generation*

Under the measurement conditions where the Rayleigh range (*i.e.,* axial extent of the focal spot) of the excitation beam is much smaller than the crystal thickness, SHG can be observed only from the vicinity of the microcrystal surface due to phase matching conditions leading to destructive interference even from a SHG-active bulk sample (37). Still, the observed stripe pattern of the SHG signal that appears in Fig. 6(b) may be attributed to the ferroelectric domains of MAPbI$_3$ as seen after etching the crystals. Unfortunately this stripe pattern was observed only in some of the measured samples (see Fig. S11). This may be due to the sample preparation procedure (whereby the fragments are not always faceted along the axes exhibiting domains), but could also indicate that the source of the SHG is in the uneven surface of the crystal. The signal dependence on polarization, which revealed a dipolar-like pattern similar in orientation (Fig. 6(c)) across the measured area regardless of the exact position, is in agreement with the presence of ferroelectric domains. The reason is that in each one of the domains the crystal orientation, which determines the SHG polarization response, is similar throughout the scanned area.

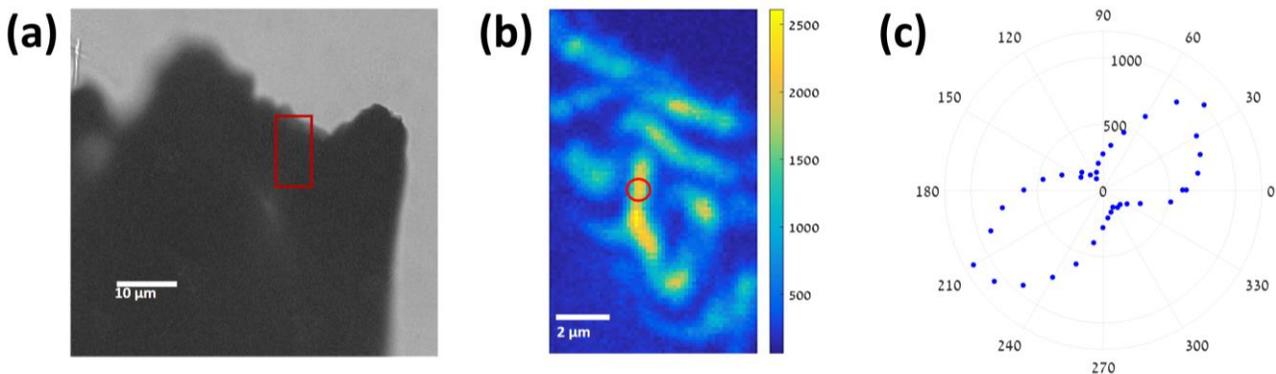

**Fig. 6: *Second harming generation (SHG) response*;** *(a) Bright field transmission image of a crystal fragment. (b) A scan of the SHG signal from an area of a crystal fragment, marked approximately by the red box; the patchy nature of the SHG signal may indicate the ferroelectric domains. The color bar is in photon counts per second (CPS). (c) A polar plot of the SHG signal from the marked point in (b), which reveals a dipolar-like pattern from CPS vs. polarization angle of the excitation laser*

Our results do not agree with a recent report on SHG in MAPbI$_3$ where no evidence of SHG activity was found in polycrystalline samples (24). We will now explain why our results do not necessarily contradict their experimental findings. In the presence of multiple ferroelectric domains, destructive interference between SHG, scattered from domains with inverted polarity, is likely to lead to extremely low SHG signals, observed in such an ensemble measurement. To be able to observe SHG in such a case, we need to *optically excite a single domain* that is larger than the scanned pixel – a case that easily can be found by scratching a mm-sized single crystal.

## Summary and conclusions

We have shown multiple experimental evidence that proves that the tetragonal phase of MAPbI$_3$ *is ferroelectric*. It behaves as a ferroelectric bulk material (*i.e.*, polarization inversion under an external electric field – Fig. 3) and it expresses all the necessary ferroelectricity-related features: lack of an inversion symmetry (proven by existence of SHG – Fig. 6), spontaneous polarization (proven by showing existing pyroelectricity – Fig. 2), and presence of polar domains (as seen after chemical etching – Fig. 5).

By analyzing the imaginary part of the permittivity, $\varepsilon_{im}$, we find the coercive and saturation fields to be ~ 0.6 kV·cm$^{-1}$ and ~ 2 kV·cm$^{-1}$, respectively. The saturation change in polarization was found to be ~ 7 µC·cm$^{-2}$ – a value that is very close to one of the theoretical estimates (~ 8 µC·cm$^{-2}$) (23). We note that, to obtain the changes in the polarization, we used the response of $\varepsilon_{im}$ instead of that of $\varepsilon_{re}$, which is uncommon and, in fact, we are not aware of the use of such analysis before. The reason to do so, is, though, very clear: in a leaky dielectric (a semiconductor) as MAPbI$_3$, where the imaginary (energy-dissipative) part dominates over the real (energy-conservative) part, the dielectric response will be mostly pronounced by the loss-related permittivity, *i.e.*, $\varepsilon_{im}$.

Besides the electric response, ferroelectricity is being expressed also structurally by creating periodically ordered polar domains. We see these polar domains after etching with acetone or from analyzing the pyroelectric decay. Both give a similar result. As common in other ferroelectric (as well as ferromagnetic) materials, in MAPbI$_3$ the relation between domain periodicity, $\omega$, and the crystal thickness, $D$, also follows the Landau-Lifshitz-Kittle model scaling relation as $\omega \sim D^{0.6}$. We find that crystal preparation may influence the $\omega$-$D$ relation – possibly as a result of difference in doping, *i.e.*, electrical conductivity. Based on the etching and periodic structure, the domains seem to form a head-to-head orientation.

The existing pyroelectricity and SHG are two expressions of the crystallographic nature of the material with no applied external fields. The results confirm that the material has a non-centrosymmetric structure and is polar, which will hopefully put an end to the confusion in the literature regarding the space-group of this tetragonal HaP: it is *I4cm* and not *I4/mcm*. It is still interesting to know if such symmetry is common to all organic-inorganic tetragonal symmetries, or something specific to MAPbI$_3$. While this question about the symmetry is proven by the existence of pyroelectricity, the recent report on the absence of SHG in MAPbI$_3$ (24) gave reason to doubt, and, thus, we show clear evidence for SHG and explain why no SHG was found in ref. (24).

How will ferroelectricity affect the operation of photovoltaic cells, based on MAPbI$_3$? To answer this question we need to take into account the leaky nature of MAPbI$_3$, which will prevent the build-up of an internal electric field to help separate electrons and holes due to spontaneous polarization. However, within a single domain this effect might possibly play a role by reducing intra-domain charge-recombination. Still, there is a quite a road ahead to find evidence in favor of such a mechanism (of interest as we want to explain the low electron-hole recombination rate). Our finding that etching reveals a domain-like structure, implies a high density of charge-carriers at the domain walls, which might fit the idea of 'ferroelectric highways' (54).

We stress that semiconducting ferroelectric materials are much less abundantly studied than normal insulating ferroelectric materials, on which our knowledge of ferroelectricity is built on. The leaky nature of SC-ferroelectric materials, as exemplified here, together with the highly dynamic nature of the MAPbI$_3$ lattice (especially at RT – Fig. 4) made the task of proving its ferroelectric nature very challenging. Analysis of the most dominant dielectric response, probing mm-sized single crystals with a defined orientation, and probing a small fraction of this crystal when looking for a SHG response has been proven here to be a good general strategy.

## Experimental

*Precursor solution*: 714 mg of methylammonium iodide, MAI (Dyesol) and 2070 mg of PbI$_2$ (99.999%, Sigma-Aldrich) were mixed with 15 ml acetonitrile (HPLC grade, ≥99.7%, BioLab). For a 1:2 molar ratio of PbI$_2$:MAI the amount of Pb was 1035 mg. The solution was sonicated for 5 min and then mixed with 1.5 ml of hydroiodic acid (57% in H$_2$O, distilled, stabilized with hypophospho­rous acid ≤1.5%, 99.95%, Sigma-Aldrich).

*Crystal Growth*: MAPbI$_3$ single crystals were grown at room temperature (below the tetragonal→ cubic phase transition temper­ature, 330 K; Fig. S12) using the slow Vapor Saturation of an Antisolvent (VSA) method (38, 8). A glass vial with the precursor solution was placed in a second, larger glass vial containing an anti-solvent - ethyl acetate (AR, ≥99.7%, BioLab) or diethylether (AR, BioLab). We used equal volumes of the solution and the anti-solvent. The lid of the inner vial was loosened to control the anti-solvent evaporation into the precursor solution. The lid of the outer vial was thoroughly sealed to prevent water penetration into the system and leakage of the solvent and/or anti-solvent vapor. Millimeter sized crystals were taken out of the system, dried (with a 'Kimwipe', Kimtech Science Brand) and stored in a silica-filled dry box.

We present results, based on 10 different crystals of MAPbI$_3$. Three out of the ten were used for pyroelectricity experiments, 2 for ferroelectric domain switching experiments, 9 for exposure of periodically ordered domains and one for SHG measurements. Other supporting results, mainly from attempted ferroelectric domain switching that led to complete destruction of the sample, are not presented. In a few cases we used crystals that were grown from modified precursor solutions (changing the amount of the stabilizer (hypophosphorous acid), normally present in HI (39), or varying the precursor ratio) to check if the polar nature of MAPbI$_3$ depends on these parameters. Apart from some quantitative differences (see Fig. 5(c) and 3S), no qualitative differences were observed.

### Crystallographic identification and preparation:

For investigation of a polar dielectric crystal, identification of the polar crystallographic orientation is required. Following powder X-ray diffraction of pulverized crystals, which confirmed the grown crystals to be MAPbI$_3$ (40), we determined the polar direction of a crystal, using specular diffraction from the surface of a single crystal (see Section 3 and Fig. S1 in the Supporting Information). After selecting crystals of a rectangular-prism shape, of a few mm size, their crystallographic orientation was identified to be the {110} and {001} planes (see Fig. S1). The planes, orthogonal to the polar direction in a *I4cm* space group (to which polar MAPbI$_3$ should belong) (24), are the {001} ones. The {110} planes, which are orthogonal to {001}, are perpendicular to non-polar directions of the crystal. To maximize the polar nature of an analyzed single crystal, all the crystals were grown at room temperature (RT), which is *below* the cubic-to-tetragonal phase transition temperature of MAPbI$_3$ at 330 K (see Fig. S12). The reason is that cooling down through the phase transition might create orthogonally oriented polar domains, which would complicate data analysis.

After planar identification, the crystals were cleaved to ~ 0.5-1 mm thick plates with an area of ~ 6-9 mm$^2$.

*Crystal preparation for electrical characterizations*: The expression of both pyroelectricity and polarization change in ferroelec­tric hysteresis loops is a function of the net dipole change in the material. Therefore, probing the change across the potentially polar direction (*i.e.,* <001>) should, conceptually, give the most significant expression of polarization change, induced by changing the temperature, *i.e.*, pyroelectricity, and for ferroelectric domain switching, by applying bias. For this reason, after identifying the crystal faces by x-ray diffraction (see Fig. S1), the crystal was cleaved along the {001} planes. For the cleaving, a fine home-made alignment station and a sharp blade were used (for visualization see Fig. 7(a)). Cleaving along the {001} planes was very clean, resulting in a highly reflective fresh surface. Cleaving along the {110} planes, on the other hand, always resulted in a jagged surface. This is being an additional way of identifying the crystal's orientation. Further discussion is brought in the 'Results and Discussion' section.

After cleaving, Carbon electrodes were deposited (using a conductive carbon paint; SPI) on two parallel {001} planes (see Fig. 7(b)). To make electrical contacts to the measuring circuit, one face of a crystal was placed on top of a copper plate (set as the *Low* terminal), while the opposite face was connected to a copper wire (set as the *High* terminal).

*Pyroelectricity*: A periodic temperature change (Chynoweth) method, was used to measure pyroelectricity (41–43). A general scheme of the setup (heat-generation and pyroelectric current measurement) is visualized in Fig. 7(c). A sample was placed in a thermally isolated Faraday cage through which $N_2$ gas flowed. The temperature of the sample's surrounding was controlled *via* a constant flow of cooled $N_2$ gas and electrical resistance heaters. The temperature was measured by a K-type thermocouple coupled to the copper plate, on which the crystals were mounted.

Periodic heating was achieved by pulsing a 1470 nm IR-laser, triggered by a function generator. A sharp rectangular pulse (rise time << 0.01 ms) was set to 15.2 ms with an overall duty cycle of 12.5%. The area on which the laser hit the crystal (~ 3-8 $mm^{-2}$) received ~ 1.1 $W \cdot cm^{-2}$. Under these conditions, the overall heating that can be expected is up to 3 K. The thermally stimulated electrical response (TSER) was amplified through a current to voltage amplifier and was read by an oscilloscope synchronized to the function-generator, which triggered the laser. The currents are normalized to the thickness of the crystal and the area of the electrodes, to get a result that allows comparison between the scanned crystals (see Section 2 in the Supporting Information) (42).

*Ferroelectricity*: To determine the dielectric nature of $MAPbI_3$, we measured the permittivity, $\varepsilon$, as function of external DC bias, $E_{DC}$, using a high-resolution impedance analyzer (Alfa, Novocontrol, Germany) at 2,222 Hz with $V_{AC}$=0.1 V (see Fig. 7(d)). The electrical leads (carbon electrodes), connector architecture (Cu plate/crystal/Cu wire) and temperature control were identical to those used for the pyroelectricity setup. The choice of frequency (2,222 Hz) was made to avoid the response from the carbon contacts, that could appear at higher frequencies (~ MHz), or contributions from ion migration or external discharge at lower frequencies (~ few Hz) (23).

The polarization, *P*, is calculated (following Eq. 1) by integrating $\varepsilon_{im}$ over an increment of $E_{DC}$ (the reasoning for this is discussed in the 'Results and Discussion' section). Therefore, to extract the correct change in polarization, the integration is done over the *absolute* incremental change in the applied bias, multiplied by the sign of the polarity of the applied bias. Due to degradation of the crystal under high bias, the reported results are from a single (first) $P$-$E_{DC}$ scan, at each temperature.

*Ferroelectric domain-size evaluation*: Two approaches were used to estimate the domain size (or 'domain periodicity', $\omega$). The first is etching a crystal in acetone for 120 sec under continuous stirring of the solvent around the crystal at RT. The crystal was photographed using a light microscope (Zeiss AXIO with a DO3THINK/ DS CB Y300E-H digital camera). The second approach was via the pyroelectric decay. At the amplifications of $10^7$-$10^9$ that were used, the discharge time of the amplifier is ~ 0.02-1 msec. Estimates of polar domain thickness were made, using the signals collected at the lower amplification ($10^7$) to avoid distortion of the signal's profile (which was found to be of the order of 0.5-5 ms). Further clarification of domain-size estimates from the pyroelectric decay signal, can be found in Section 2 in the Supporting Information.

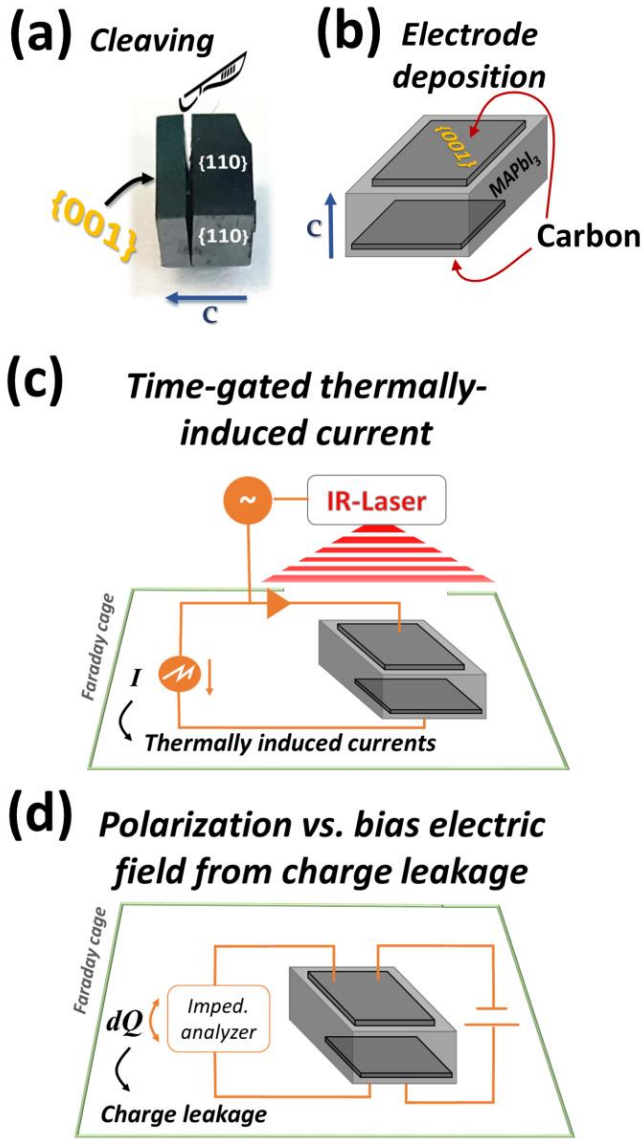

*Fig. 7: **Illustrations for sample preparation and dielectric-nature measurements**; (a) Photograph of a MAPbI$_3$ crystal (#3), cleaved along its {001} planes after identifying its polar direction. (b) Illustration of the locations where the carbon electrodes were deposited. (c) General scheme of pyroelectricity generation and measurement setup. The circle with the '~' symbol indicates a function generator. A circle with the 'W' symbol indicates an oscilloscope. A triangle symbol indicts a current to voltage amplifier. (d) General scheme of the polarization measurements under external bias.*

<u>Second Harmonic Generation (SHG) measurements</u>: 100 fs pulses @ 800 nm at an 80 MHz repetition rate from a Ti:Sapphire laser (Coherent, Chameleon Ultra II) were down-converted, using a synchronously pumped OPO (Coherent, Chameleon Compact OPO), to get the 1100 nm laser wavelength, desired for excitation. The laser was directed into a microscope (Zeiss Axiovert 200 inverted microscope) and focused using an oil immersion objective (Zeiss Plan Apochromat X63 NA 1.4). A clean-up polarizer and a rotation stage with a half-wave plate were placed before the microscope to control the excitation polarization. The spectrally filtered, (Semrock, 561/40, Thorlabs KG3 heat glass) signal was epi-detected (*i.e.*, excitation and collection were from the same side of the crystal), coupled to a multimode fiber and detected by a single-photon avalanche photodiode (Perkin-Elmer, SPCM), which was connected to a time-correlated single-photon counting (TCSPC) system (Picoquant HydraHarp 400). Samples were placed onto a No. 1 cover slip, yielding some crystal fragments, prepared by scraping a crystal's {110} face, with dimensions of ~ 10-100 µm. SHG images were formed by scanning the excitation spot along the side lying on the cover slip. The exact crystallographic orientation of each scanned face was not analyzed.

**Acknowledgements**

Y.R. thanks Dr. Isai Feldman for guidance with XRD measurements, Dr. Omer Yaffe, Igal Levin and Arava Zohar (all from the Weizmann Institute) for fruitful discussions. We thank Prof. V. M. Fridkin (Shubnikov Inst. Cryst., Moscow) for suggesting the low temperature approach to find a ferroelectric loop, and to Prof. S. Trolier-McKinstry, (Penn.State University) for constructive remarks regarding Fig. 3. We are grateful to Dana and Yossie Hollander via the Weizmann Institute's Alternative Sustainable Energy Research Initiative, the Israel Science Foundation (to IL), to the Israel Ministry of Science's Tashtiot, Israel-China and India-Israel programs, and to the Israel National Nano-initiative, for partial support, and acknowledge the historic generosity of the Harold Perlman family. D.C. holds the Sylvia and Rowland Schaefer Chair in Energy Research.

# *Supplementary Information*

### Tetragonal CH$_3$NH$_3$PbI$_3$ Is Ferroelectric

Yevgeny Rakita[1], Omri Bar-Elli[2], Elena Meirzadeh[1], Hadar Kaslasi[1], Yagel Peleg[1], Gary Hodes[1], Igor Lubomirsky[1], Dan Oron[2], David Ehre[1,a], David Cahen[1,a]

[1] *Materials and Interfaces Dept., Weizmann Institute of Science, 7610001, Rehovot, Israel*

[2] *Physics of Complex Systems Dept., Weizmann Institute of Science, 7610001, Rehovot, Israel*

[a] Electronic mail: david.ehre@weizmann.ac.il ;  david.cahen@weizmann.ac.il

**Table of Contents:**

**Section 1**: *Origins of polarity in perovskites and its detection*

Ionic displacement due to distortion from the ideal cubic perovskite *ABX₃* structure is the origin for local or long-range polar nature of perovskites. Perovskites are stacked as corner-sharing [*BX₆*] octahedrons forming cuboctahedral sites occupied by an *A* cation. A common origin for structural distortion in perovskites is the mismatch between the ionic volume of the different species which make up a unit cell – a mismatch that can be quantified by a ratio of ionic radii, known as Goldschmidt tolerance factor (1). Such imperfection leads to tilting and axial rotation of the relatively rigid corner-sharing octahedra, which lowers the symmetry of the structure (2, 3). As the *A* cation is usually spherically symmetric, tilt and rotation of the [*BX₆*] octahedron are not enough to form a dipole in a unit cell. Ionic displacements inside the octahedron, however, can lead to polar or anti-polar unit-cells (4, 5). One known chemical origin for ionic displacement in oxide perovskites is the unequal probability of the $O^{2-}$ valence electrons to interact with the orbitals of the non-centrosymmetric *B* valence electrons. This lack of symmetry may create an ionic displacement. In $BaTiO_3$, for example, the interaction between the non-centrosymmetric valence $Ti^{4+}$ *d*-orbital electrons and $O^{2-}$ *p*-orbital electrons cause a small ionic displacement of the $Ti^{4+}$ within the [$TiO_6$] octahedron, which leads to a polar unit cell and to ferroelectricity (6).

In HaPs, however, the valence orbitals of the *B* group are known (at least for the *p*-block Pb, Sn or Ge) to be mainly *s*-orbitals, which are centrosymmetric (*e.g. 6s* in $MAPbX_3$).(7, 8) Therefore, the chemical origin for spontaneous polarization in $MAPbX_3$ is hypothesized to be sustained by the polar organic group (7). This idea is supported by results from neutron diffraction (9–11), which showed that free rotation of $MA^+$ create an effectively cubic system (space group: $Pm\bar{3}m$). As the system is cooled down, vibrationally degenerate soft modes of the $MA^+$ and/or [$PbX_6$] octahedron are being condensed to a lower degree of degeneracy, meaning that they keep vibrating but with less degenerate modes. This forces the system to take on tetragonal symmetry and by further cooling, a lower degree of degeneracy forces it to become orthorhombic (9).

The reason that a space group notation is omitted after 'tetragonal' and 'orthorhombic' is due to a confusion that pervades the scientific literature. Based on x-ray- and neutron- diffraction techniques, the tetragonal phase of $MAPbI_3$ (<330 K) was ascribed to a non-polar symmetry space group, *I4/mcm*, as well as to a polar one, *I4cm* (12). Using Aleksandrov notations for tilt/displacement notation for perovskite-like systems (5), *I4/mcm* and *I4cm* have exactly the same octahedron-tilt system, (00ϕ), and the only difference is that *I4cm* is also polar along the *c*-direction (having $a = b \neq c$, with a notation for polarity of (00p); cf. Fig. Si for illustration). Following this logic, $MAPbI_3$ can also possess a *Cmca* space group if the polar displacement along the *c*-axis is anti-polar, (00a). The debate regarding the orthorhombic phase (<165 K) is also between a centrosymmetric - *Pnma*, (ϕϕψ)/(000) - and a polar – *Pna2₁* one with (ϕϕψ)/(0p0)). For MAPbBr₃ the debate concerns only the orthorhombic phase (<149 K), similar to what is the case for MAPbI₃, where the tetragonal phase (<236 K) was always noted as *I4/mcm* (9).

The interpretation of data from diffraction-based techniques depends on the user's choice of the model to which the diffraction pattern should fit. Small deviations, such as between *I4/mcm* and *I4cm*, can rarely be clearly distinguished, causing disagreements in the literature in oxide perovskites, but not only there (13, 14).

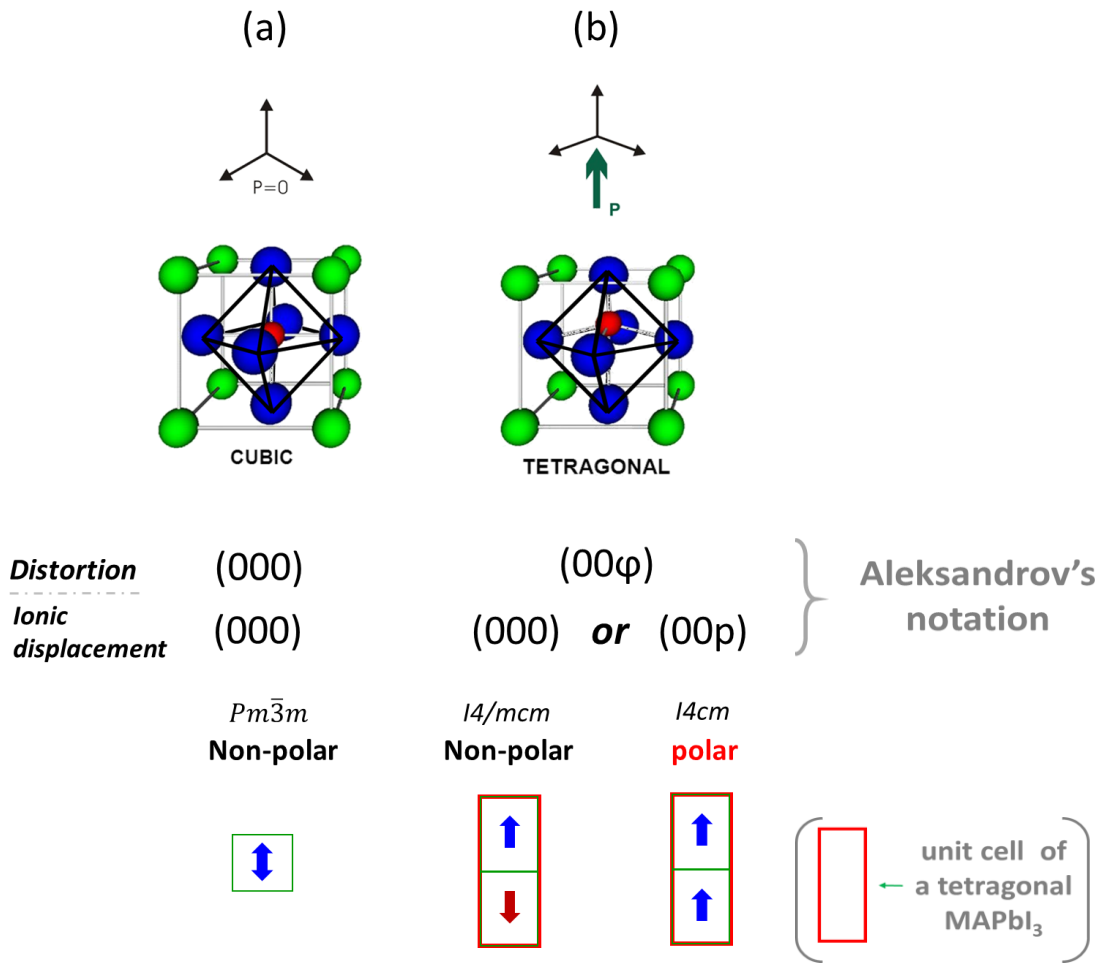

**Fig. Si:** Illustrations of the (a) cubic and (b) tetragonal phases relevant to the MAPbI₃ perovskite structure. The tetragonal symmetry can fit to two possible space groups that are very hard to distinguish, using diffraction methods. The difference will result in a material with a different dielectric nature.

Several methods can be used instead of diffraction methods, based on the relations between dielectric properties and bulk symmetry (see Fig. Sii), to get an idea about the symmetry-dictated dielectric nature of a material:

(1) *Polarization (P) vs. electric field (E)* scan hysteresis loops are the most direct evidence for ferroelectricity. As mentioned in the main text, high electrical conductivity and/or low stability will limit the interpretation of results using capacitance-based measurement systems, such as 'Sawyer Tower'-based methods (15).

(2) *Pyroelectricity*, will provide evidence for spontaneous polarization in the bulk material. Before concluding this, alternative explanations for a thermally stimulated electrical response (TSER) need to be excluded.

(3) Piezo-response force microscopy can give a partial understanding regarding the polar nature of the material('s surface). This technique can reveal ferro-*elasticity*(16) and piezoelectricity (17), which indicate a non-centrosymmetric space group of a material. However, the method, which has been quite popular in halide perovskite research, cannot directly determine the polar nature of the material.

**(4)** Second harmonic generation (SHG) will occur whenever a material lacks a centre inversion symmetry in at least one crystallographic direction, meaning that a material is non-centrosymmetric (18).

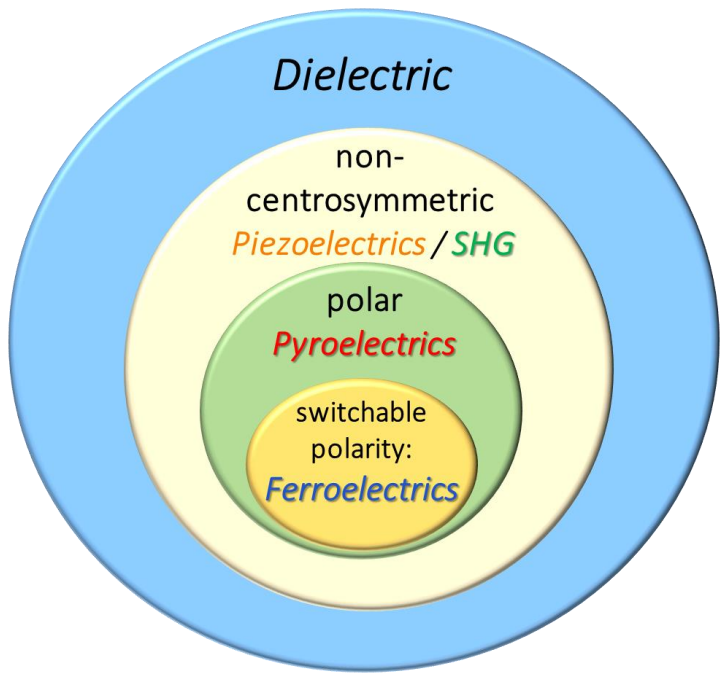

**Fig. S ii:** Schematic illustration of the progression from more general to more specific dielectric properties and techniques to identify them. For example, the scheme illustrates that if a material is ferroelectric it must be polar and non-centrosymmetric.

**Section 2:** *Pyroelectric current response and calculation of domain period, ω*:

The pyroelectric current, $I_p$, is given by the following relation (19):

S1) $$I_p = I_0 \cdot erf\left(\frac{\omega}{\sqrt{4 \cdot \delta \cdot t}}\right)$$

S2) $$I_0 = \frac{A \cdot F_d \cdot \alpha}{C_V \cdot (d + \omega)}$$

where $\omega$ is the size of the polar domain, $\delta$ [cm$^2$·s$^{-1}$] is the thermal diffusion coefficient, $t$ [s] is the time of during the laser pulse, $I_0$ [A] is the current at $t = 0$, $C_V$ [J·K$^{-1}$·cm$^{-3}$] is the thermal capacitance per unit volume, $\alpha$ [C·K$^{-1}$·cm$^{-2}$] is the pyroelectric coefficient, $F_d$ [W·cm$^{-2}$] is the heat flux at the surface, $A$ [cm$^2$] the electrode area and $d$ [cm] is the sample thickness. Since $I_p$ depends on $d$ and $A$, it is customary to present a normalized current, $i_p$, instead of $I_p$:

S3) $$i_p = \frac{d}{A} \cdot I_p.$$

From the pyroelectric current, we can estimate the polar domain size as follows. Using the *erf* table and equation S1 we can write that at half-maximum current ($i=0.5i_0$):

S4) $$\frac{\omega}{\sqrt{4 \cdot \delta \cdot t_{HM}}} = 0.48$$

where $t_{HM}$ is the decay time from the maximum current to the half-maximum current after subtracting the system delay time (20 μs for our system). Now, if we assume that $\delta = 6 \times 10^{-3}$ cm$^2$·s$^{-1}$ for MAPbI$_3$ (20) and estimate from Fig. 4(b) that $t_{HM} = 140$ sec, we find, using equation S2, $\omega \sim 9$ μm.

### Section 3: *Crystallographic identification from specular x-ray diffraction*

X-ray diffraction is performed in the Bragg-Brentano reflection geometry using a TTRAX III (Rigaku, Japan) a $\theta$-$\theta$ diffractometer, equipped with a rotating Cu anode operating at 50 kV and 200 mA and with a scintillation detector aligned at the diffracted beam. For general crystallization product identification crystals were pulverized and powder diffraction of the resulting sample was performed. After confirming that the result conformed to the known diffractogram of MAPbI₃, the natively exposed faces of a crystal were aligned parallel to the specular plane of the scan to identify the {001} plane, which corresponds to the polar direction in *I4cm* symmetry. To distinguish between the slight differences of {001} and {110} crystallographic orientations, a high $2\theta$ angle was scanned. This is done to allow diffraction from the well separated n=4 multiples of the (002) and (110) diffractions at $2\theta$=58.2° and 58.8°, respectively. As some crystals showed both diffractions (most likely due to twinning, which is very probable for tetragonal symmetry), we used further only crystals with highly oriented {001} planes for our studies. Fig. S1 shows an example of an actual diffraction result from different spontaneously exposed planes of an MAPbI₃ crystal.

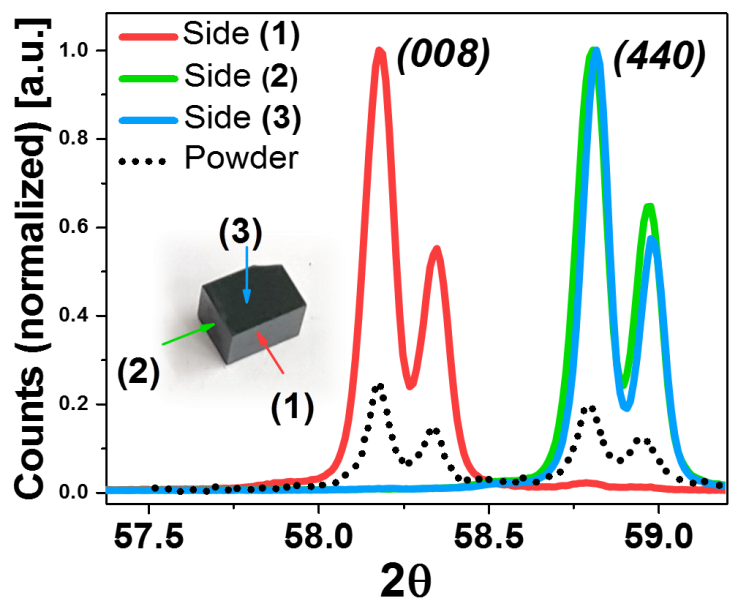

**Fig. S 1:** (a) Specular x-ray diffraction from crystal #3 (as shown in the inset). Three well-developed sides of the crystal were scanned after aligning the crystal face parallel to the specular plane. The $2\theta$ angle was set to allow diffraction from the n=4 reflections of the (002) and (110) diffractions. We found that 'sides 1' corresponded to the {001} faces and the other two sides corresponded to the {110} faces, as expected from the tetragonal symmetry of MAPbI₃. The two adjacent peaks result from the Cu Kα₁ and Kα₂.

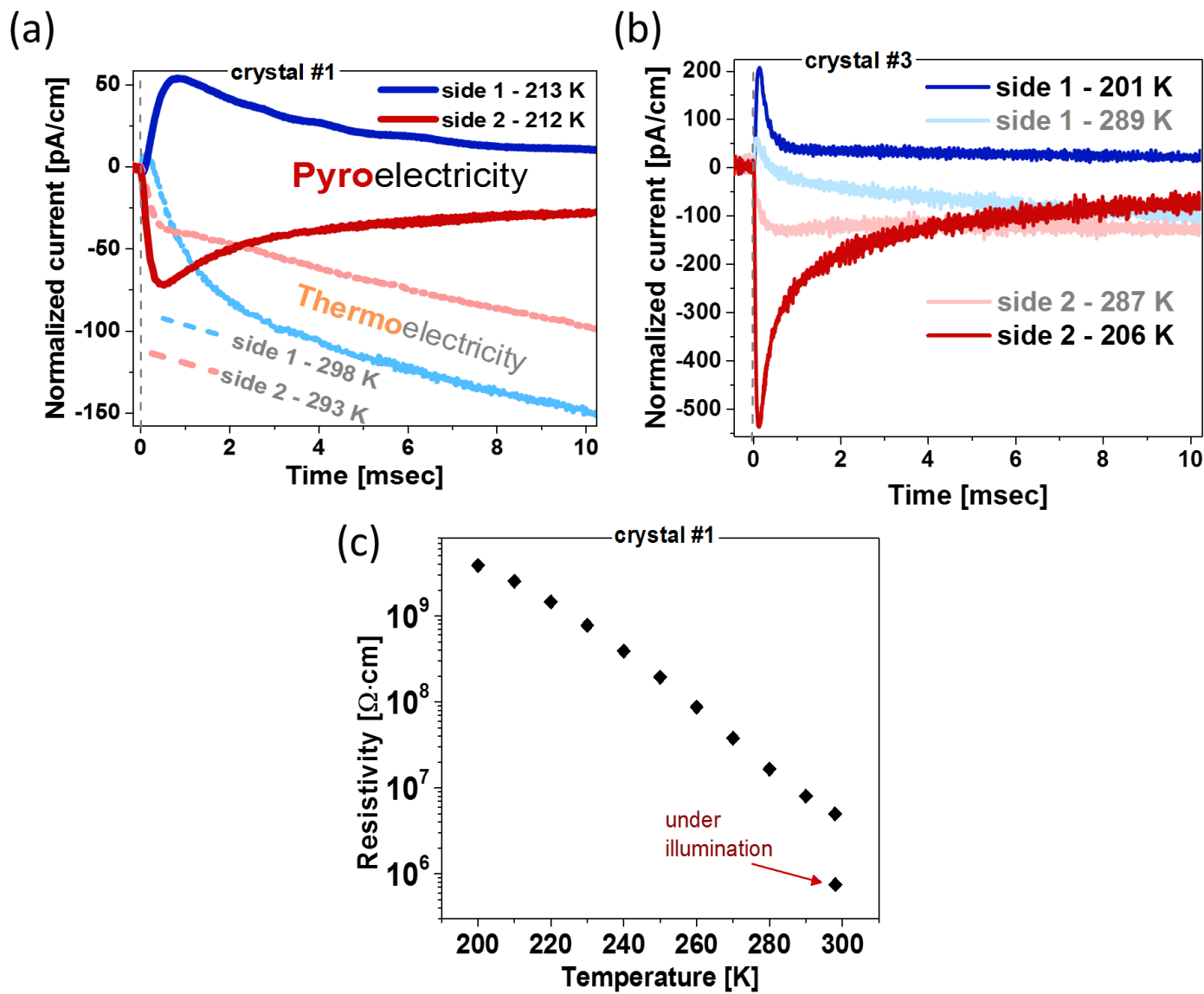

**Fig. S 2:** TSER response at RT (pale-colored plots) and low temperatures (deep-colored plots) of two crystals, in addition to the one, used for the results shown in Fig. 2 in the main text: (a) crystal #1: $PbI_2$:MAI molar ratio = 1:1; crystal morphology – hexagonal (diamond-like) prism. (b) Crystal #3: $PbI_2$:MAI molar ratio = 1:2; crystal morphology – rectangular prism. We used ethyl acetate as an anti-solvent for the growth of these two crystals. The polar nature of the $PbI_2$:MAI - 1:2 crystal is more pronounced, but is not absent in the 1:1 crystal. For crystal #2 (shown in the main text) we used diethyl ether as an anti-solvent and a $PbI_2$ to MAI ration of 1:2.

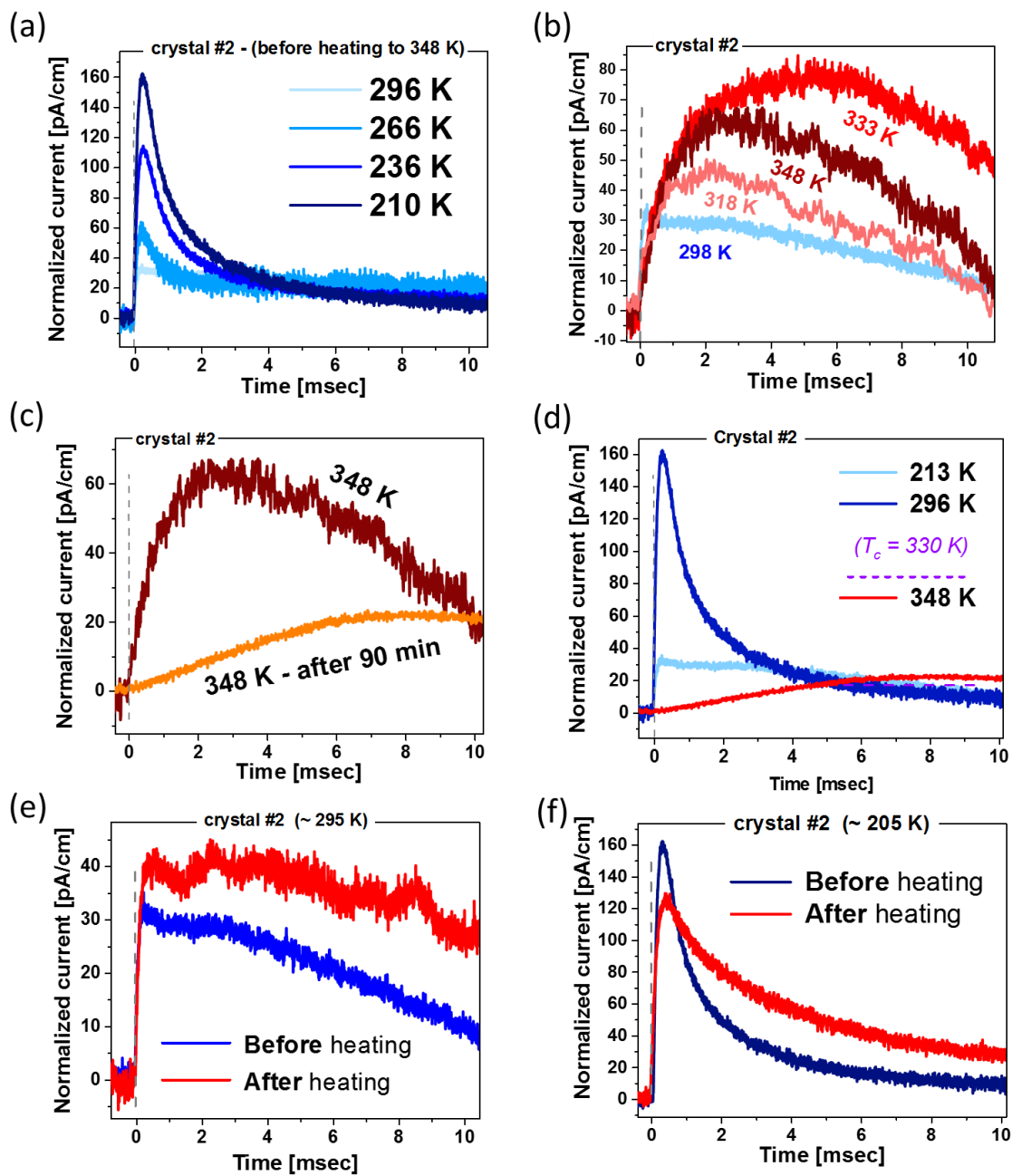

**Fig. S 3:** TSER dependence on environmental temperature as evidence for the pyroelectric nature of the TSER signal. (a) TSER upon cooling from RT; (b) TSER upon heating from RT; (c) TSER after 10 minutes and after 100 minutes from the moment the temperature reached 348 K; (d) comparison between the low, RT and high (>330 K) temperature TSER signals; (e) TSER at RT before and after heating to 348 K for 100 min; (d) TSER at low temperature before and after heating to 348 K for 100 min.

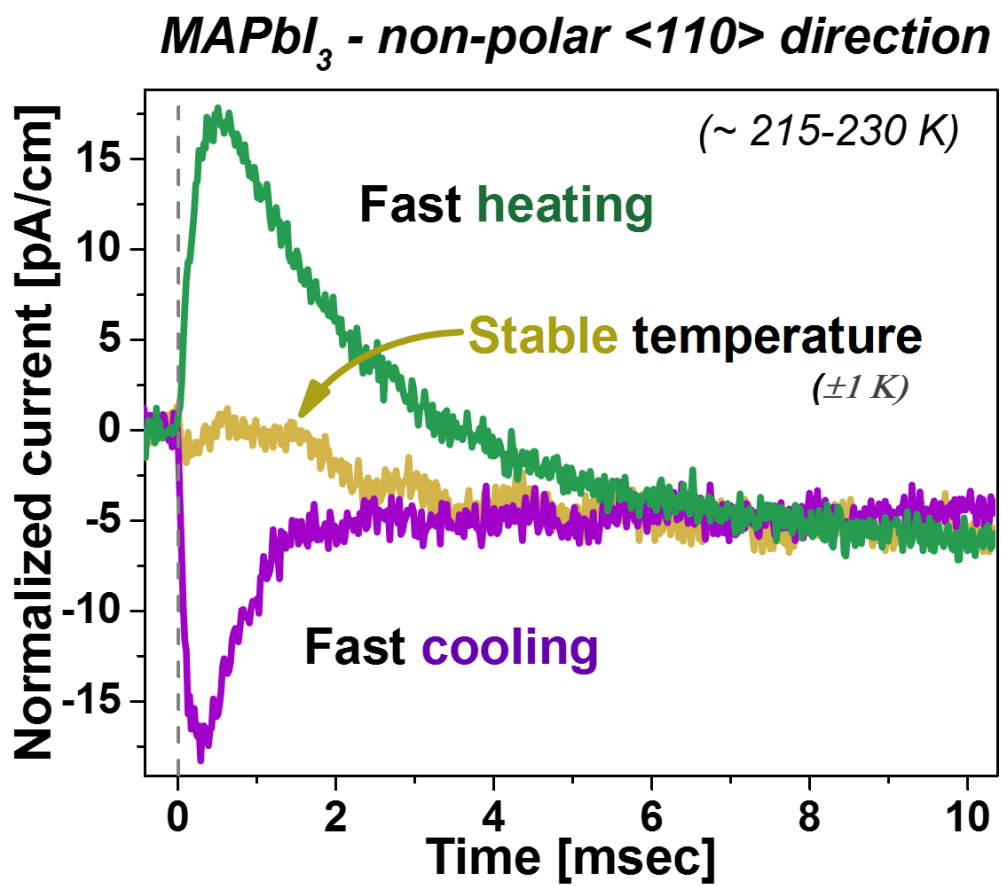

**Fig. S 4**: TSER at LT along the {110} plane of a MAPbI$_3$ crystal. Upon fast heating or cooling the peak current can be reversed, which strongly indicates these TSER signals are related to flexoelectric polarization (creation of dipole due to strong inhomogeneous thermal expansion along the crystal). If the temperature is stable (±1 K) the flexoelectric currents disappear and the small remaining current is, most likely, related to residual thermoelectric currents.

# MAPbBr$_3$

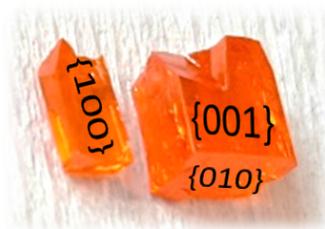

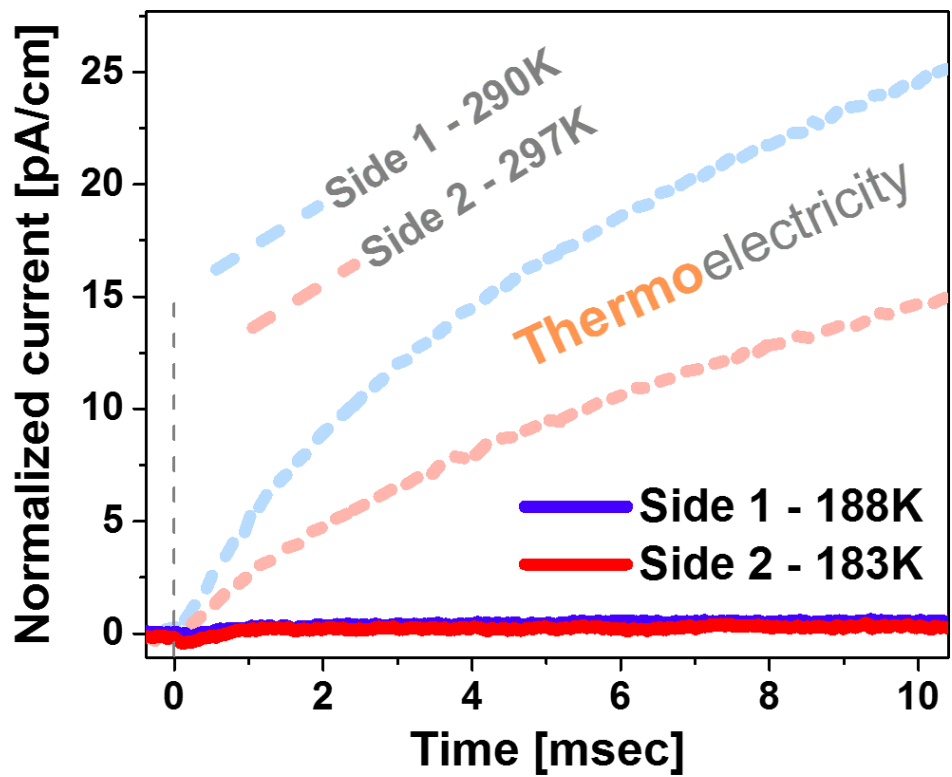

**Fig. S 5**: TSER of MAPbBr$_3$ along one of the {100} directions showing only thermoelectric currents (pale colored plots) and zero response at ~185 K deep-colored plots – a temperature well below its cubic to tetragonal phase transition temperature (@ 235 K and well above the tetragonal to orthorhombic one, 145 K). The opposite thermoelectric current sign of the MAPbBr$_3$ - positive - crystal to that of MAPbI$_3$ - negative – is explained by the different Fermi level position with respect to the valence and conduction bands: TSER indicates that MAPbBr$_3$ isa p-type and MAPbI$_3$ is n-type.

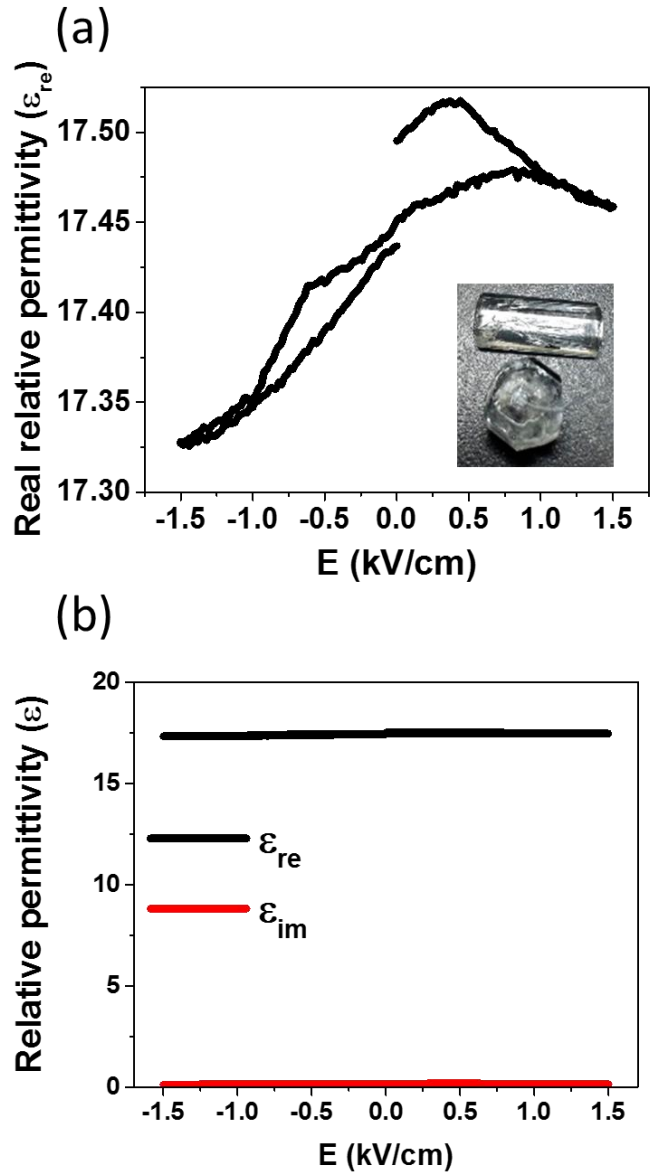

**Fig. S 6:** Dielectric response of a crystal of Rochelle salt (Potassium Sodium Tartrate, $KNa(OC(O)C(OH))_2\cdot4H_2O$) cleaved across its polar direction to a ~1 mm thick plate. The dielectric response is measured with an impedance analyzer (at 122 kHz and 1 $V_{AC}$) under a bias electric field (see experimental section, Fig. 1(d)). The measuring temperature was 273 K, where the polarization response to a bias electric field, $E_{DC}$ is maximal.(15) (a) The real part of the relative permittivity, $\varepsilon_{re}$, as function of $E_{DC}$. The change in the dielectric response is small (fraction of a percent), but still shows a very similar profile to that of $Pb(Zr_xTi_{1-x})O_3$, PZT, in which the change in dielectric response can be more than 100% (21). (b) Comparison of the real, $\varepsilon_{re}$, and the imaginary (lossy) part, $\varepsilon_{im}$, of the relative permittivity. It is clearly seen that $\varepsilon_{im}$ can be neglected for this material.

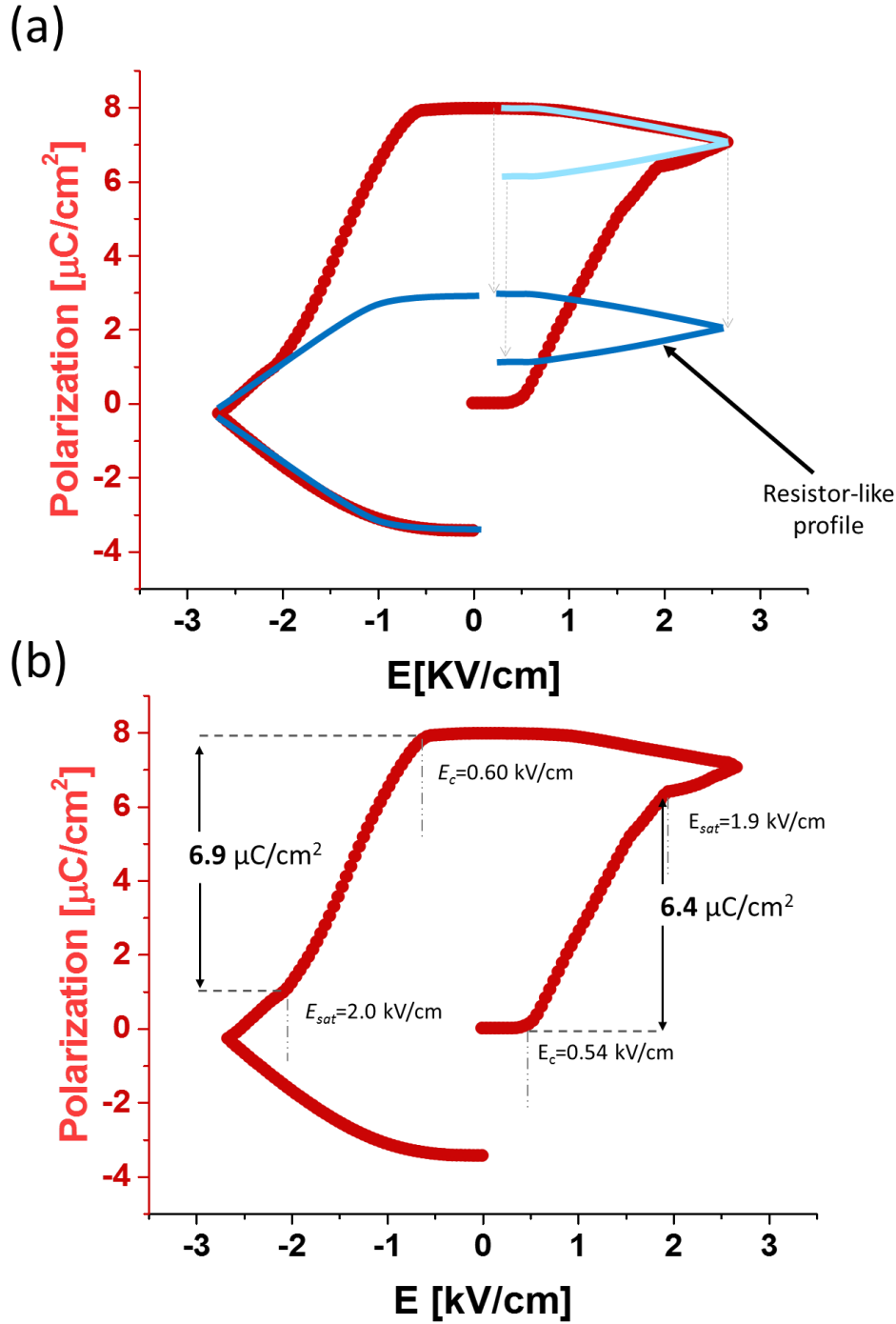

**Fig. S 7:** (a) Qualitative deconvolution of the hysteresis loop, shown in Fig. 3(b) illustrating that the loop is composed of a lossy part (i.e., resistor-like polarization; blue) and a ferroelectric part (sharp transitions due to ferroelectric polarization). (b) Analysis of the ferroelectric polarization part. The average between the positive and negative bias of the coercive field ($E_c$), saturation field ($E_{sat}$) and saturation change in polarization ($P_{sat}$) show similar values on both the positive and negative parts of the loop, indicating true ferroelectric switching. The positive and the negative sides of the loop are not equivalent due to asymmetric contacts or some irreversible degradation due to the high applied electric field.

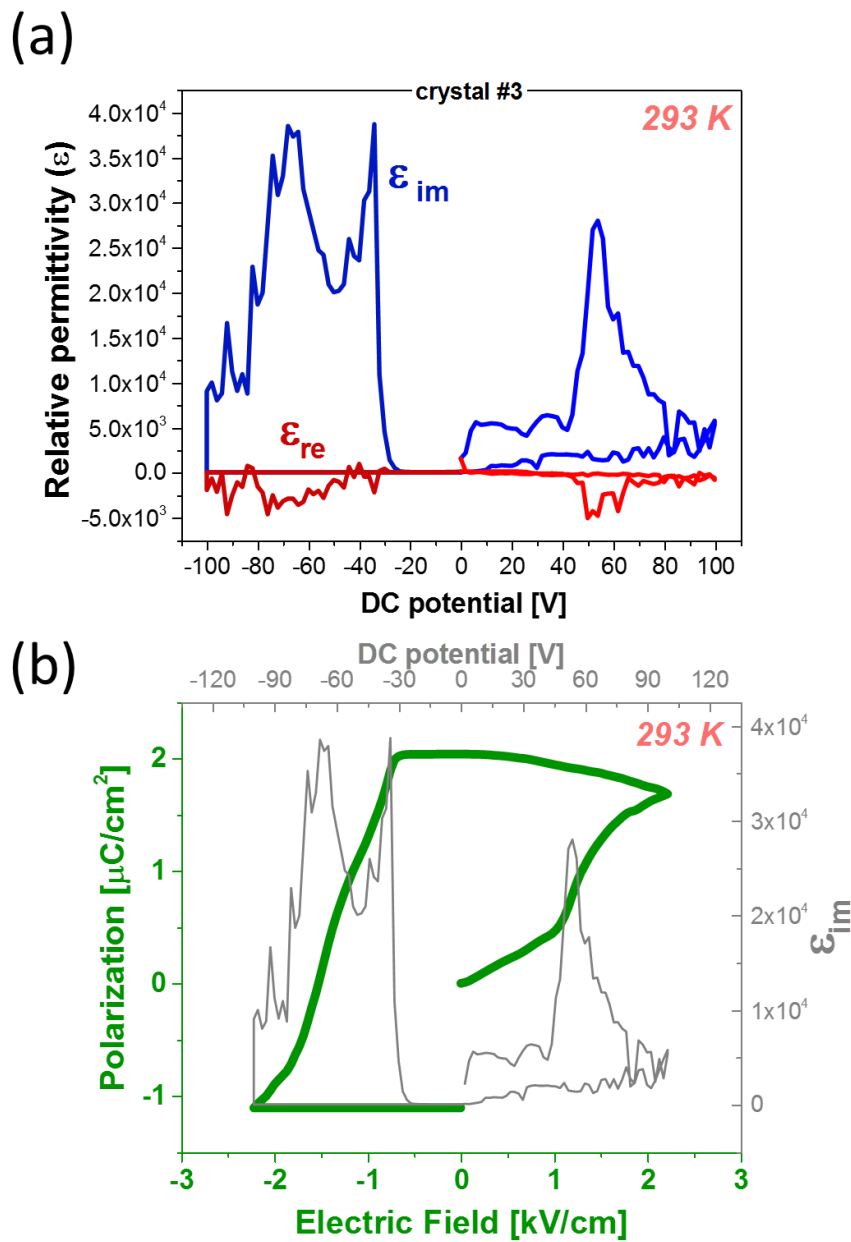

**Fig. S 8:** (a) Dielectric response at RT of a MAPbI₃ crystals along its <001> direction 2222 Hz (V$_{AC}$ =0.1 V) as a function of an applied bias, $E_{DC}$. The $\varepsilon_{im}$ - $E_{DC}$ scan was done after the scan at LT. Similarly, to the response at LT, the RT response shows a dominant imaginary (lossy) part of the dielectric response. (b) P-E hysteresis loop obtained from integration of $\varepsilon_{im}$ over $E_{DC}$ (see Eq. 1). The greater (negative) values in $\varepsilon_{re}$ can be related to electrochemical irreversible degradation due to the applied electric field (and voltage), that eventually cause severe damage to the sample (see lower left part of the loop). This degradation does not allow a fair analysis of the loop (positive part - before the severe damage) as ferroelectric switching.

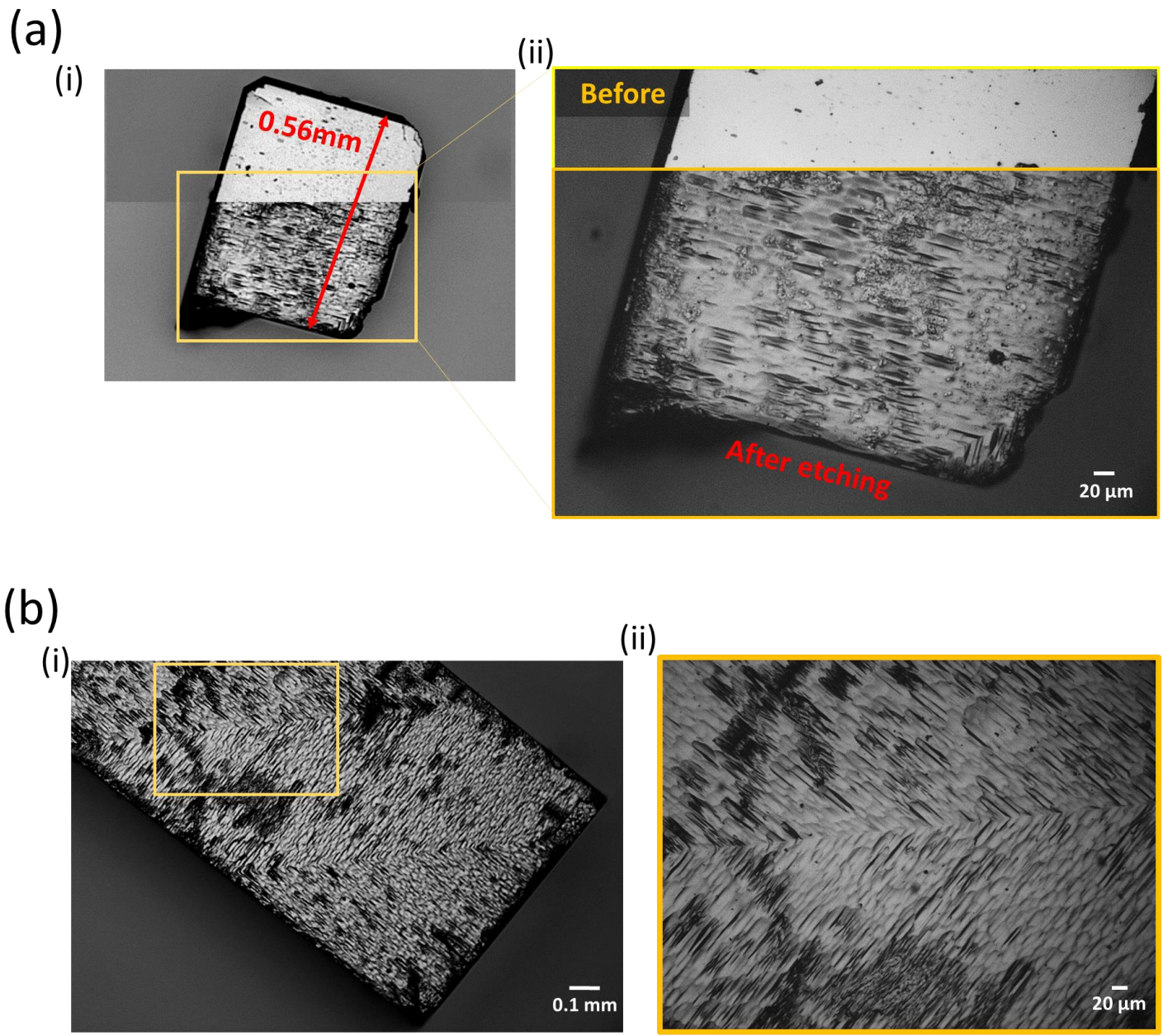

**Fig. S 9:** Bright field image from a light microscope of a crystal before and after etching in acetone for 120 sec. (a) and (b) are crystals grown in a 1:1 PbI$_2$ to MAI solution using diethyl ether as an anti-solvent. (a) is found to be a single crystal while (b) is a crystal with two twin boundaries. At such a twin boundary the polar domains are orthogonal to each other. (ii) are magnified view of the yellow rectangles marked in (i).

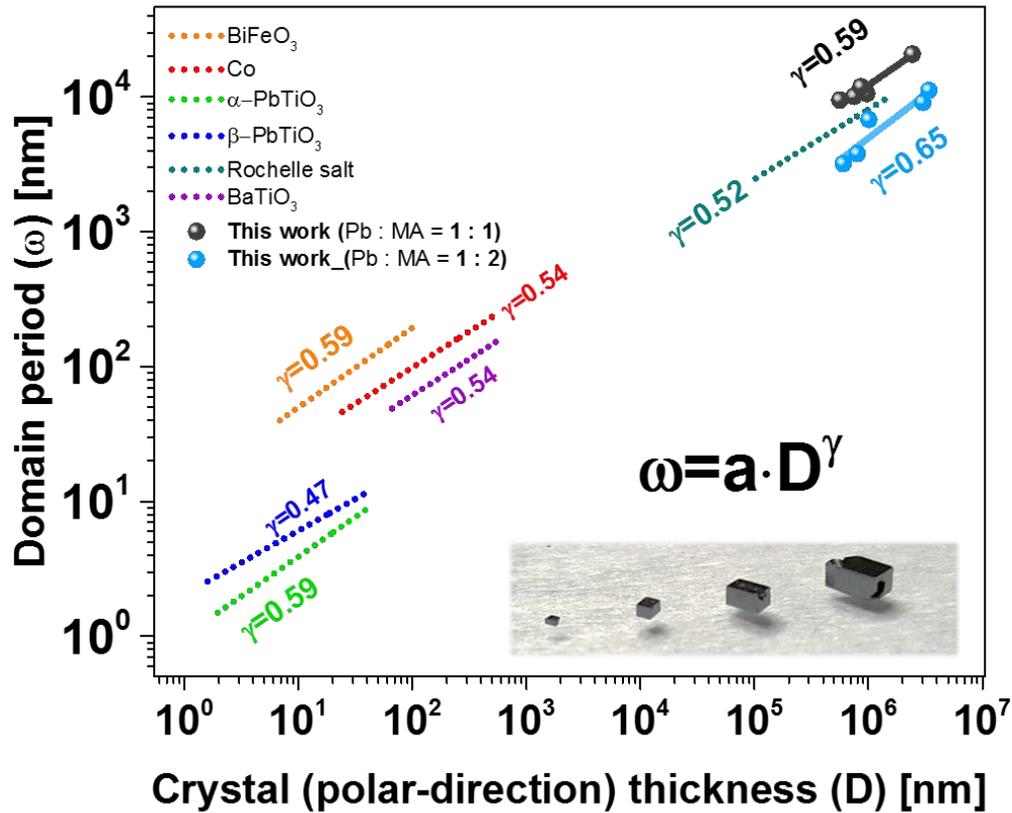

**Fig. S 10:** Domain periodicity with respect to the crystal thickness in the polar direction. The data are an average obtained from several microscope images after etching in acetone for ~120 sec. The values of $\gamma$ are obtained by fitting to a $\omega = a \cdot D^{\gamma}$ function. The fitted lines for the other ferroic materials are taken from refs. (22, 23). These are introduced to compare with our results, showing good agreement with the global scaling law between bulk sample dimensions and domain periodicity.

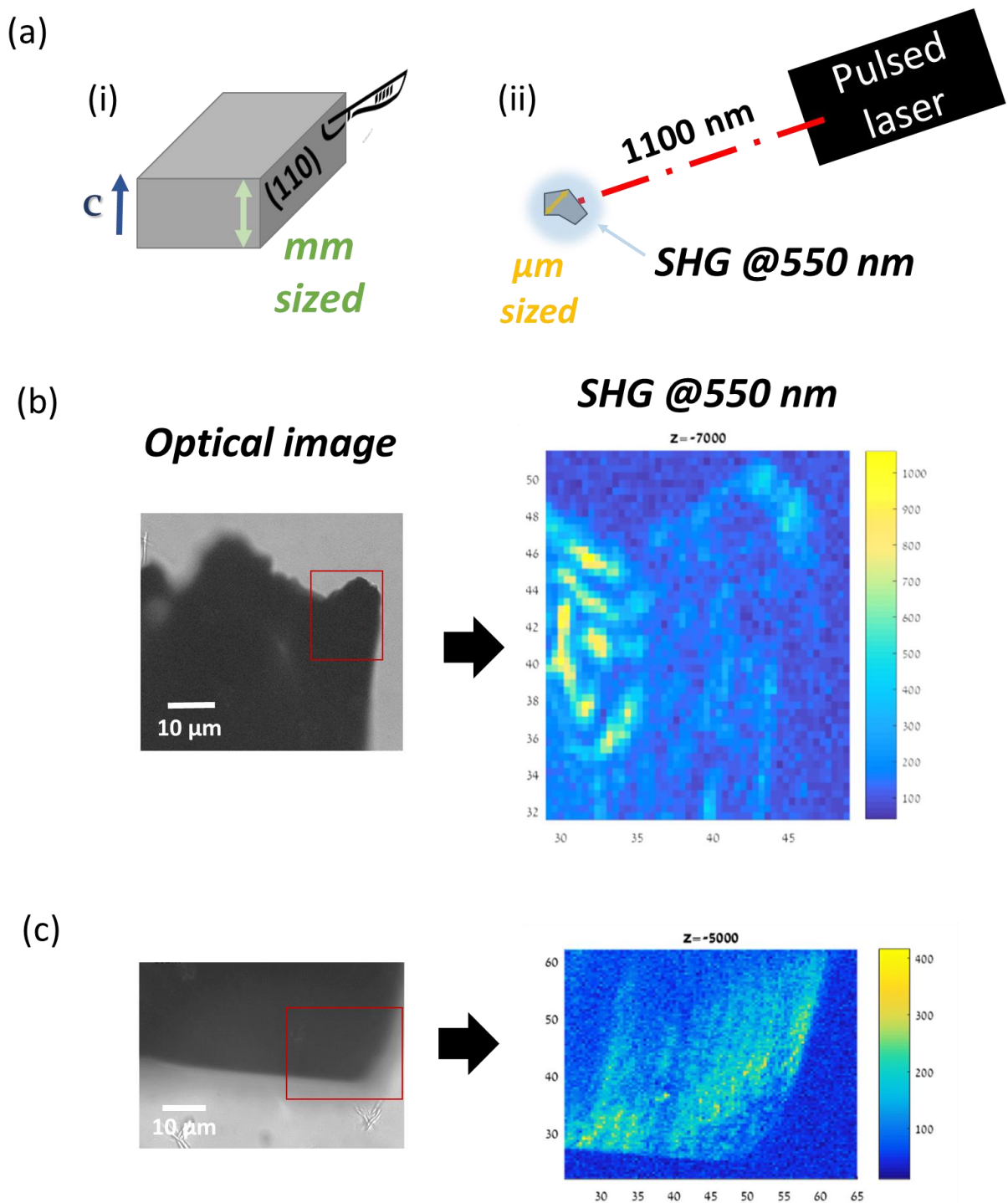

**Fig. S 11:** (a) I illustration of (i) the way a sample was prepared for the SHG detection experiment and (ii) its measurement. (b) and (c) are examples of results from different pieces of such samples (*cf.* also Fig. 6 in main text).

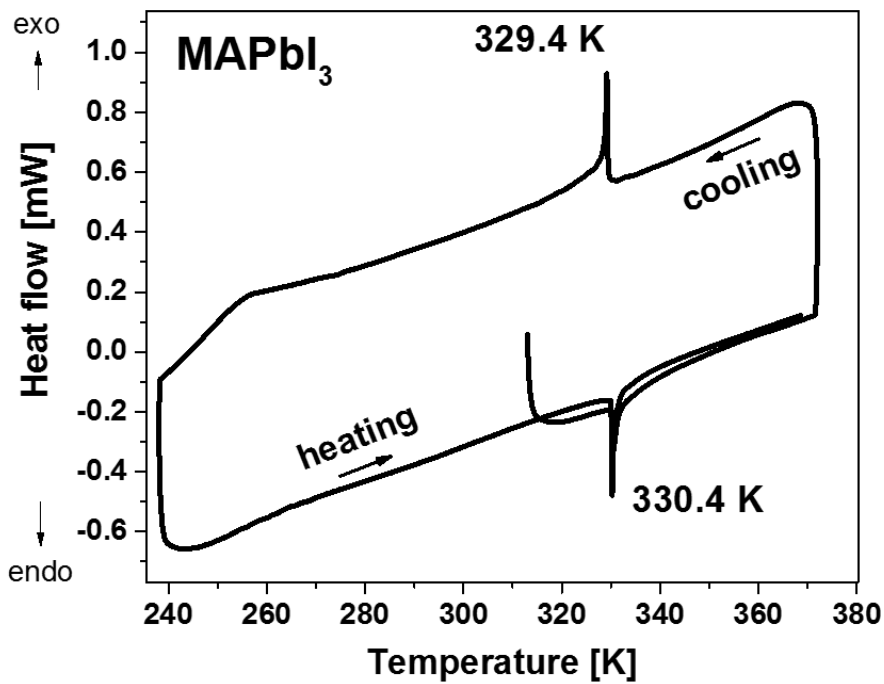

**Fig. S 12:** Differential scanning calorimetry (DSC) of 0.3-1 mm crystals grown below the tetragonal-to-cubic phase transition as described in the experimental section. DSC was carried in a Q200 model from "TA instruments" using hermetically sealed aluminum pans under $N_2$ flow at a heating rate of 10 °C/min and cooling rate of 5 °C/min.